\documentclass[lettersize,journal]{IEEEtran}
\usepackage{amsmath,amsfonts}
\usepackage{algorithmic}
\usepackage{algorithm}
\usepackage{array}
\usepackage[caption=false,font=footnotesize,labelfont=rm,textfont=rm]{subfig}
\usepackage{textcomp}
\usepackage{xcolor}

\usepackage{stfloats}
\usepackage{url}
\usepackage{verbatim}
\usepackage{graphicx}

\usepackage{cite}
\usepackage{bm}
\usepackage{cite}
\usepackage{amssymb}
\usepackage{float} 
\usepackage{caption}
\usepackage{textcomp}
\usepackage{upgreek} 
\usepackage{amsthm}
\usepackage{subfig}		
\usepackage{cuted}
\makeatletter
\renewcommand{\maketag@@@}[1]{\hbox{\m@th\normalsize\normalfont#1}}%
\makeatother
\usepackage{booktabs}
\usepackage{hyperref}
\usepackage{tabularx}  
\begin{document}
\title{A Harmony Composition-Inspired Tensor Modalization Method for Near-Field IRS Channel Estimation}

\author{Wenzhou Cao, Yashuai Cao, 
Tiejun~Lv,~\IEEEmembership{Senior Member,~IEEE},
and~Jie~Zeng, \IEEEmembership{Senior Member,~IEEE}

\thanks{Manuscript received 13 April 2025; revised 15 October 2025 and 23 January 2026; accepted 7 March 2026. This paper was supported by the National Natural Science Foundation of China under No. 62271068. (\emph{corresponding author: Tiejun Lv}.)}
\thanks{Wenzhou Cao and Tiejun Lv are with the School of Information and Communication Engineering, Beijing University of Posts and Telecommunications (BUPT), Beijing 100876, China (e-mail: \{cwzhou9511, lvtiejun\}@bupt.edu.cn).}
\thanks{Yashuai Cao is with the School of Intelligence Science and Technology, University of Science and Technology Beijing, Beijing 100083, China (e-mail: caoys@ustb.edu.cn).}
\thanks{Jie Zeng is with the School of Cyberspace Science and Technology, Beijing Institute of Technology, Beijing 100081, China (e-mail: zengjie@bit.edu.cn).}
}

\maketitle

\begin{abstract}
Intelligent reflecting surfaces (IRSs) are poised to revolutionize next-generation wireless communication systems by enhancing channel quality and spectrum efficiency through advanced wave manipulation. However, extremely large-scale IRS {(XL-IRS)} deployments face significant challenges in channel estimation due to multiplicative path loss and near-field (NF) effects, where spherical wavefronts couple distance and angle parameters. Existing polar-domain codebook-based compressive sensing methods for NF channel estimation suffer from low accuracy and high complexity, caused by the need for high-resolution grids of both distance and angle parameters. To address this, we propose a harmonic processing-inspired channel estimation framework for NF {XL-IRS} systems by leveraging tensor modalization to decouple channel parameters. Drawing an analogy to musical harmonic analysis, our approach decomposes the high-dimensional NF channel tensor into independent factor matrices, modeled as ``chords," representing distance and angle parameters.
Through harmonic analysis-inspired distance parameter decoupling, we design a compact, distance-dependent codebook that enables high-resolution NF channel parameter estimation. This approach significantly reduces the codebook size compared to polar-domain methods.
{Then, we} derive the Cram\'{e}r-Rao lower bound (CRLB) to evaluate the estimators. Finally, simulation results show an 8.5 dB improvement in normalized mean square error (NMSE) compared to conventional methods, underscoring its low complexity and high accuracy.
\end{abstract}


\section{Introduction}
\IEEEPARstart{I}{ntelligent} reflecting surfaces (IRSs) have been regarded as a key technology for future sixth-generation (6G) mobile communications. Through advanced wave engineering capabilities, IRS can offer appealing advantages in solving signal blockage, enhancing channel quality, and optimizing spectrum utilization~\cite{r1,r2,r3}. However, it is reported that IRSs suffer from severe performance degradation due to the \emph{multiplicative path loss effect}~\cite{r4}. To address this issue, a high-density array configuration of the IRS, namely, extremely large-scale IRS {(XL-IRS)}, is desired in the next-generation wireless networks. This involves integrating massive low-cost and passive reflective elements to geometrically expand the electromagnetic (EM) wave control region meanwhile reaping the large aperture gains~\cite{r5}. 
Despite these benefits, the practical implementation of {XL-IRSs} is hampered by two critical challenges: \textit{i)} prohibitive complexity of high-dimensional channel estimation, and \textit{ii)} near-field (NF) effect~\cite{r6} arisen from the extended Rayleigh distance. In the NF radiation region, EM waves exhibit spherical wavefront characteristics. This means the channel path difference is affected not only by the propagation angles but also by the propagation distances between the IRS and user equipments (UEs)~\cite{r7, r8}. 
As a result, developing an efficient {XL-IRS} channel estimation scheme by accounting for NF effect becomes imperative for the commercial deployment of 6G.

Extensive efforts have been conducted on channel estimation methods for NF IRS-assisted wireless systems~\cite{r9,r10,rWu,r12,r13}. The authors of~\cite{r9} proposed to utilize several active reflective units for NF channel estimation. However, such IRS configurations with active reflective units inevitably introduce additional power consumption and signal processing complexity. 
In \cite{r10}, a NF IRS channel estimation was transformed into a compressed sensing (CS) problem, where the polar domain orthogonal matching pursuit algorithm (PSOMP) was developed.
To reduce the correlation between polar-domain codebook entries, an improved polar-domain codebook was designed~\cite{rWu}, where the orthogonal least squares (OLS) algorithm was utilized to solve the sparse channel estimation.
Through polar domain transformation, the authors of~\cite{r12} constructed a three-dimensional multi-measurement vector CS framework for NF IRS channel estimation and a low-complexity estimation strategy was proposed. 
In~\cite{r13}, the NF channel was converted into a sparse representation in the polar domain. Then, the authors modeled channel parameters as Bernoulli-Gaussian distributed random variables, and proposed an expectation-maximization (EM) algorithm for NF channel estimation. 
However, the aforementioned channel estimation methods mainly rely on the polar domain codebook for sparse transformation. The channel estimation performance heavily depends on the sampling grid resolution of the polar domain codebook. High-resolution codebook can reap high estimation accuracy but requires a large codebook size, especially for NF polar-domain codebook. To evaluate the impact of the codebook size on the NF channel estimation performance, {Fig. 1 represent the maximum correlation value between the NF channel parameters of the polar domain codebook and the real NF channel parameters. A larger polar domain codebook size indicates higher resolution of the codebook, and higher resolution means that the polar domain codebook is closer to the true channel parameters, resulting in a higher percentage.}

\begin{figure}[!t]
\centering
\includegraphics[width=2.9in]{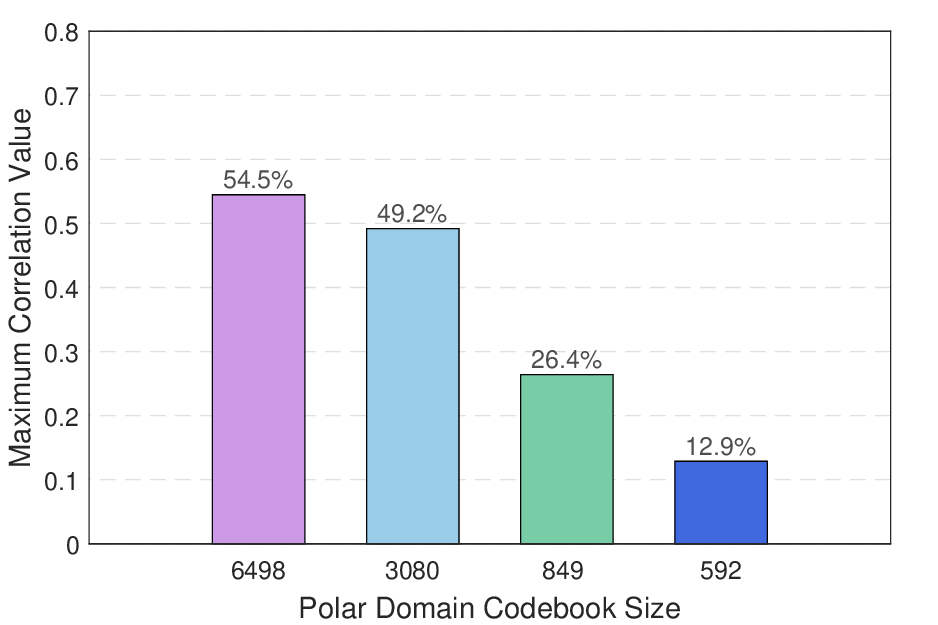}
\caption{Maximum correlation value between varying-sized polar domain codebooks and true values.}
\label{fig_1}
\end{figure}

In addition, a variety of advanced channel estimation techniques tailored for IRS scenarios have been examined~\cite{r14, r15, r16, r17, r18}. Specifically, a novel anchor-assisted channel estimation scheme was presented in~\cite{r14} to reduce the pilot overhead. However, this scheme requires considerable feedback and is not suitable for sparse multipath channels. In \cite{r15} and \cite{r16}, a low-rank matrix completion-based channel estimation algorithm was designed for the low-rank IRS channels. Based on the sparsity of millimeter-wave (mmWave) channels, the channel estimation problem was transformed into sparse signal recovery problem~\cite{r17, r18}. {The authors construct an observation signal reception tensor model using the base station-IRS channel, the IRS-UE combined pilot channel, and the IRS phase change matrix, and treat the IRS reflection elements as tensor ranks ~\cite{a1, a2, a3, a4, a5}. For the factor matrices of the constructed tensor model, the authors employ the alternating least squares (ALS) algorithm and the vector approximate message passing (VAMP) scheme to estimate the channel factor matrices. However, these methods involve large-scale matrix operations and iterative message updates, which further increase the overall computational complexity.} 

    Unlike far-field plane-wave channel, where the array response depends solely on angular parameters, the NF spherical-wave channel response is governed by both angular and distance. This dependency stems from path differences across array elements relative to a reference point. Mathematically, the NF array response is fundamentally defined by the path difference between array elements; the angle parameter emerges merely as a derivative representation of the spatial relationship between the reference distance and the coordinate axes. Consequently, the distance and angle parameters exhibit a dominant-subordinate relationship. However, traditional polar-domain methods fail to exploit this dominant-subordinate relationship. They merely perform a joint grid searchs that suffer from high computational complexity. Inspired by the dominant-subordinate relationship between the root and auxiliary tones in harmony theory, this work designs a dedicated tensor-based distance-angle parameter decoupling. By first estimating the dominant distance and then constructing a reduced-dimensional codebook for angle extraction, our approach fundamentally diverges from existing NF tensor models to achieve efficient and accurate parameter estimation.
    The main contributions of this paper are summarized as follows:
\begin{itemize}

\item { We propose a harmonic processing-inspired NF IRS channel estimation framework based on tensor modalization. This framework consists of three key steps: \textit{i)} chord construction-based tensor canonical polyadic decomposition (CPD) for structural modeling; \textit{ii)} harmonic analysis-inspired distance parameter decoupling and factor matrix estimation; and \textit{iii)} chord progression-guided angle parameter extraction. The proposed framework enables a high-accuracy and low-complexity estimation of uplink channel parameters in NF IRS communications by circumventing the curse of NF codebook resolution.}

\item {Similar to the way identical root notes appear across different chord configurations in harmonic structures, we exploit the inherent structure of shared distance parameters between the NF channel delay term and the IRS array response term, representing these terms as the delay factor matrix (tonic chord) and IRS array factor matrix (dominant chord), respectively. The user-side angular component is modeled as the user angle factor matrix (subdominant chord). This tensor factorization approach transforms the complex high-dimensional channel estimation problem into the optimization of sparse path parameters. Drawing from harmonic analysis, which distinguishes between different chords, we apply singular value decomposition (SVD) to the mode-1 unfolding of the observed signal tensor. Leveraging the Vandermonde structure inherent in the delay factor matrix, we convert the SVD results into eigenvalue decomposition (EVD) form to estimate the distance parameters. The resulting delay factor matrix then enables effective estimation of both the IRS array and user angle factor matrices.}

\item {Drawing inspiration from the tonic chord's function in directing harmonic progressions, the estimated delay factor matrix is employed to create a distance-dependent yet dimension-reduced codebook, in contrast to the polar-domain approach. Through correlation detection with the estimated IRS array or user angle factor matrix, the angle-of-arrival (AoA) and angle-of-departure (AoD) parameters are effectively extracted.}

\item {We derived the Cramér-Rao lower bound (CRLB) for the NF tensor channel and the associated channel parameters. Since the distance parameters are tightly coupled in the tensor factor matrices, we applied the chain rule to separate them from the different factor matrices, completing the CRLB calculation for the distance parameters. We also analyzed the relationship between the CRLB and the signal-to-noise ratio (SNR). Simulation results show that our approach outperforms the polar compressed sensing (CS) method in terms of estimation accuracy and is closer to the CRLB.}
\end{itemize}

The remainder of the paper is structured as follows: Section~\ref{II} discusses the uplink channel model for NF IRS communications. The harmonic processing-inspired channel parameter estimation algorithm is presented in Section~\ref{III}. The CRLB of the proposed channel parameter estimator is derived in Section~\ref{IV}. Simulation results are presented in Section~\ref{V}, followed by conclusions in Section~\ref{VI}.

\textit{Notations}: Bold-face upper- and lower- cases indicate matrices and vectors, respectively. ${\left(  \cdot \right)^\mathsf{T}}$, ${\left(\cdot\right)^{\mathsf{H}}}$, $\left (\cdot \right) ^*$, ${\left(\cdot  \right)^{-1}}$, and ${\left(\cdot\right)^\dag }$stand for matrix transpose, conjugate transpose, conjugate, inverse and pseudoinverse, respectively. $\left\|\cdot\right\|$ and ${\left\|\cdot\right\|_F}$ represent the 2-norm and Frobenius-norm, respectively. $\otimes$, $\odot$, and $\circ$ denote the Kronecker product, Khatri-Rao product and outer product, respectively.

\begin{figure}[t]
\centering
\includegraphics[width=3.1in]{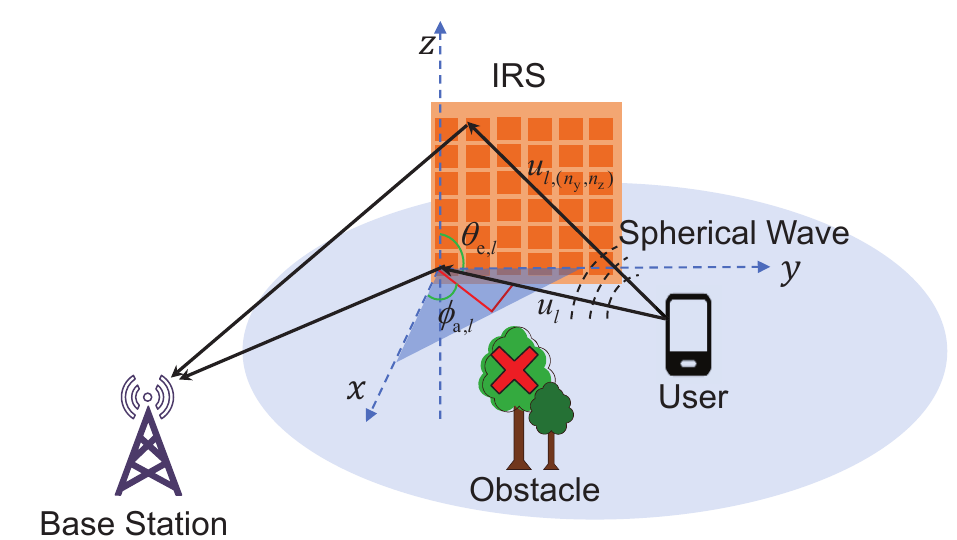}
\caption{NF IRS-assisted system model.}
\label{fig_2}
\end{figure}

\section{NF IRS System Model}\label{II}
As shown in Fig.~\ref{fig_2}, we consider an uplink orthogonal frequency-division multiplexing (OFDM) transmission system, where the base station (BS) and the UE are equipped with the uniform linear array (ULA). The BS and UE arrays comprise ${N_{\rm{b}}}$ and ${N_{\rm{t}}}$ elements, respectively. 
The IRS is configured with a uniform planar array (UPA) comprising ${N_{\rm{r}}} = {N_{\rm{y}}}{N_{\rm{z}}}$ elements, arranged in ${N_{\rm{y}}}$ rows and ${N_{\rm{z}}}$ columns. The phase shift matrix of the IRS is denoted by ${{\bm{\Omega}}} = {\mathrm{diag}}\left( \bf{v} \right) = {\rm{diag}}\left( {{e^{j{\omega _1}}},{e^{j{\omega _2}}}, \ldots, {e^{j{\omega _{{N_{\rm{r}}}}}}}} \right)\in \mathbb{C}^{N_{\rm r} \times N_{\rm r}}$, where $\mathbf{v} = \left [ {{e^{j{\omega_1}}},{e^{j{\omega _2}}}, \ldots, {e^{j{\omega _{{N_{\rm{r}}}}}}}}  \right ]\in \mathbb{C}^{ 1\times N_{\rm r} }$  and the phase shift is ${\omega _i} \in \left[ {0,2\pi } \right]$. The ${{\bf{H}}_{\rm{u}}}$ and ${{\bf{H}}_{\rm{s}}}$ represent UE-IRS and IRS-BS channels, respectively. Here, we assume that the communication link between the BS and the UE is blocked.

\subsection{Channel Model}
A single-sided NF channel is adopted, as the {XL-IRS} employs a large-aperture UPA with a significantly larger number of elements compared to the UE side. The total number of subcarriers in the communication system is ${{P_0}}$, and ${{P}}$ subcarriers are selected for CE training.  Meanwhile, our system is a narrowband system. {At the ${{p}}$-th subcarrier, the channel between the IRS and the UE is expressed as~\cite{b6}}

\begin{equation}
{{\bf{H}}_{{\rm{u,}}p}} = \sum\limits_{l = 1}^L {{\gamma _l}{e^{ - j2\pi {f_{\rm{s}}}{\tau _l}\frac{p}{{{P_0}}}}}} {{\bf{a}}_{{\rm{IRS}}}}\left( {{\theta _{{\rm{e}},l}},{\phi _{{\rm{a}},l}},{u_l}} \right){\bf{a}}_{{\rm{ue}}}^{\mathsf H}\left( {{\psi _l}} \right),
\label{eq_1}
\end{equation}
where ${{\bf{H}}_{{\rm{u,}}p}}\in \mathbb{C}^{N_{\rm r} \times N_{\rm t}}$ denotes the IRS-UE channel under the ${{p}}$-th subcarrier, $p = \left\{ {1,2, \ldots ,P} \right\}$. ${L}$ represents the number of paths between the IRS and the UE, $f_{\rm{s}}$ is the sampling frequency, ${{\gamma _l}}$ is the complex gain, ${{\theta _{{\rm{e}},l}}}$ (or ${{\phi _{{\rm{a}},l}}}$) is the elevation (or azimuth) AoA at the IRS, ${{\psi _l}}$ is the AoD at the UE, ${\tau _l} = {u_l}/c$ is the time delay, $c$ is the light speed, and ${{u_l}}$ is the distance between the reference reflective element and the scatterer (or UE) of the $l$-th path, where $l\in \left \{ 1,2,\dots ,L\right \}$. Additionally, the spherical wave array response and the plane wave array response are denoted by ${{\mathbf{a}}_{{\rm{IRS}}}}$ and ${{\mathbf{a}}_{{\rm{ue}}}}$, respectively.

In the NF spherical wave front, the NF array response is related to the angle and distance domains. Specifically, the three-dimensional (3D) coordinate is utilized to model channels, where the reference antenna coordinates of the IRS are $\left( {0,0,0} \right)$. {Then, its ${n_{\rm{y}}}$-th and ${n_{\rm{z}}}$-th antenna coordinates are $\left( {0,{n_{\rm{y}}},{n_{\rm{z}}}} \right)$, where $0 \le {n_{\rm{y}}} \le {N_{\rm{y}}}$, $0 \le {n_{\rm{z}}} \le {N_{\rm{z}}}$.
}  Based on the 3D angular information, the coordinates of the $l$-th scatterer (or UE) can be obtained as $\left( {{u_l}\sin \left( {{\theta _{{\rm{e}},l}}} \right)\cos \left( {{\phi _{{\rm{a}},l}}} \right),{u_l}\sin \left( {{\theta _{{\rm{e}},l}}} \right)\sin \left( {{\phi _{{\rm{a}},l}}} \right),{u_l}\cos \left( {{\theta _{{\rm{e}},l}}} \right)} \right)$, and the distance from the UE to the $\left( {{n_{\rm{y}}},{n_{\rm{z}}}} \right)$- IRS element is given by 
\begin{align}
{u_{l,({n_{\rm{y}}},{n_{\rm{z}}})}} =& \left( {{{\left( {u_l\sin ({{\theta _{{\rm{e}},l}}})\cos ({{\phi _{{\rm{a}},l}}})} \right)}^2}}\notag\right. 
  \\ &+ {\left( {u_l\sin ({{\theta _{{\rm{e}},l}}})\sin ({{\phi _{{\rm{a}},l}}}) 
- \left( {{n_{\rm{y}}} - 1} \right)d} \right)^2}\notag
  \\&+ {\left. {{{\left( {u_l\cos ({{\theta _{{\rm{e}},l}}}){n_{\rm{z}}}
 - \left( {{n_{\rm{z}}}
 - 1} \right)d} \right)}^2}} \right)^{1/2}}\notag\\
  = &\left( {{u_l^2} + {{\left( {{n_{\rm{y}}} - 1} \right)}^2}{d^2} + {{\left( {{n_{\rm{z}}} - 1} \right)}^2}{d^2}} \notag\right.
 \\&- 2\left( {{n_{\rm{y}}} - 1} \right)du_l\sin ({{\theta _{{\rm{e}},l}}})\sin ({{\phi _{{\rm{a}},l}}})
 \notag\\&{\ { - 2\left( {{n_{\rm{z}}} - 1} \right)du_l\cos ({{\theta _{{\rm{e}},l}}})} \Big )^{1/2}},
\end{align}
where $d$ is the antenna spacing. Since the array response is related to the wave range difference and the wave formation difference of the spherical wave front is $ \Delta{u_{l,({n_{\rm{y}}},{n_{\rm{z}}})}} = {u_{l,({n_{\rm{y}}},{n_{\rm{z}}})}} - u_l$, the array response ${{\bf{a}}_{{\rm{IRS}}}}\left( {{\theta _{{\rm{e}},l}},{\phi _{{\rm{a}},l}},{u_l}} \right)\in \mathbb{C}^{N_{\rm r} \times 1}$ in the NF is written by
\begin{equation}
\begin{array}{l}
{{\bf{a}}_{{\rm{IRS}}}}\left( {{\theta _{{\rm{e}},l}},{\phi _{{\rm{a}},l}},{u_l}} \right)= [{e^{ - j2\pi \frac{{\Delta{u_{l,1}}}}{{{\lambda _{\rm{c}}}}}}},\cdots,{e^{ - j2\pi \frac{{\Delta{u_{l,{N_{\rm{r}}}}}}}{{{\lambda _{\rm{c}}}}}}}{]^{\mathsf T}}
\end{array},
\label{eq_3}
\end{equation}
where the wavelength is ${\lambda _{\rm{c}}}$ and the carrier frequency is ${f_{\rm{c}}}$. Due to the far-field assumption at the UE side, the array response ${{\bf{a}}_{{\rm{ue}}}}\left( {{\psi _l}} \right)\in \mathbb{C}^{N_{\rm t} \times 1}$ of the plane wavefront is written by 
\begin{equation}
\begin{array}{l}
{{\bf{a}}_{{\rm{ue}}}}\left( {{\psi _l}} \right) = [{e^{ - j2\pi \frac{{d\sin {{\psi _l}}}}{{{\lambda _{\rm{c}}}}}}},\cdots,{e^{ - j2\pi \frac{{({N_{\rm{t}}} - 1)d\sin {{\psi _l}}}}{{{\lambda _{\rm{c}}}}}}}{]^{\mathsf T}}
\end{array}.
\label{eq_4}
\end{equation}

The BS-IRS channel can be characterized in the far-field form due to the large propagation distance~\cite{r20}. { In the initial deployment phase of the IRS-assisted system, its global position information can be obtained through GPS, laser ranging, or other sensor fusion technologies, thus assuming that the BS-IRS channel is known}~\cite{r21}. The BS-IRS channel ${{\bf{H}}_{{\rm{s,}}p}}\in \mathbb{C}^{N_{\rm b} \times N_{\rm r}}$ on the $p$-th subcarrier is expressed as 
\begin{equation}
{{\bf{H}}_{{\rm{s,}}p}} = \tilde \alpha {e^{ - j2\pi {f_s}\tilde \tau \frac{p}{{{P_0}}}}}{{\bf{\tilde a}}_{{\rm{B}}}}\left( {{\varphi _{\rm{B}}}} \right){{\bf{\tilde a}}}_{{\rm{IRS}}}^{\mathsf H}\left( {{\theta _{{\rm{IRS}}}},{\phi _{{\rm{IRS}}}}} \right),
\end{equation}
where $\tilde \alpha$ and $\tilde \tau $ represent the complex path gain and delay between the BS and IRS, respectively. ${\theta _{{\rm{IRS}}}}\left( {\phi _{{\rm{IRS}}}}\right)$ and ${{\varphi _{\rm{B}}}}$ represent the elevation angle (azimuth angle) at the IRS and the AoA at the BS, respectively.  {
	The vector 
    ${{\bf{\breve a}}_{{\rm{B}}}}\left( {{\varphi _{\rm{B}}}} \right) = [{e^{ - j2\pi \frac{{d\sin {{{{\varphi _{\rm{B}}}}}}}}{{{\lambda _{\rm{c}}}}}}},\cdots,{e^{ - j2\pi \frac{{({N_{\rm{b}}} - 1)d\sin {{{{\varphi _{\rm{B}}}}}}}}{{{\lambda _{\rm{c}}}}}}}{]^{\mathsf T}}$ denotes the array response at the BS and the far-field array response ${{\bf{\breve a}}}_{{\rm{IRS}}}\left( {{\theta _{{\rm{IRS}}}}, {\phi _{{\rm{IRS}}}}} \right)={{\bf{\breve a}}_y}\left( \sin {{\theta _{{\rm{IRS}}}} \sin{{\phi _{{\rm{IRS}}}}}} \right)\otimes {{\bf{\breve a}}_z}\left(\cos{{{\phi _{{\rm{IRS}}}}}}\right)$, where ${{\bf{\breve a}}_z}\left(\cos{{{\phi _{{\rm{IRS}}}}}}\right)$ and ${{\bf{\breve a}}_y}\left( \sin {{\theta _{{\rm{IRS}}}} \sin{{\phi _{{\rm{IRS}}}}}} \right)$ represent the far-field vertical and horizontal array responses, respectively, with the specific form similar to ${{\bf{\breve a}}_{{\rm{B}}}}\left( {{\varphi _{\rm{B}}}} \right)$.}


Since the static positions of the IRS and BS are easily acquired, the BS-IRS channel is assumed to be known. Therefore, we only need to complete the channel estimation between the IRS and UE, i.e., estimate the channel parameters $\left\{ {{\theta _{{\rm{e}},l}},{\phi _{{\rm{a}},l}},{\tau_l},{{\psi _l}},{\gamma _l}} \right\}_{l = 1}^{{L}}$.

\subsection{Signal Reception Model}
In the uplink channel estimation scheme employed in this paper, $T_{{\rm{a}}}$ time frames are used for uplink training, where each time frame is split into $Q$ time slots. We set $t\in \left \{ 1,2,\dots ,T_{{\rm{a}}}  \right \}$ and $q\in \left \{ 1,2,\dots ,Q \right \}$ as the indices for the time frames and time slots, respectively.
In the $q$-th time slot, the IRS reflects the signal using a fixed phase shift matrix. In the $t$-th time frame, the pilot signal ${s_p}(t)$ is sent under the $p$-th subcarrier with ${s_p}(t) = 1$, and the beamforming vector ${{\bf{x}}_p}(t) = {\boldsymbol{{\rm{F}}}_{{\rm{RF}}}}(t){{\bf{f}}_{{\rm{BB}},p}}{s_p}(t)$ is obtained after the hybrid precoding matrix and its dimension is $N_{\rm t} \times 1$, where ${{\bf{f}}_{{\rm{BB}},p}}$ denotes the digital precoding vector of the $p$-th subcarrier, assuming that ${{\bf{f}}_{{\rm{BB}},p}} ={\bf{f}_{{\rm{BB}}}}$. For all subcarriers, the analog precoding matrix can be written as ${{\boldsymbol{{{\rm{F}}}}}_{{\rm{RF}}}}$, and for simplicity of the model it is assumed that there is only one radio frequency chain at the receiver and transmitter side. {Similarly, the hybrid combination vector for all subcarriers can be expressed as ${{\bf{w}}}\in \mathbb{C}^{ N_{\rm b}\times 1 }$.
} The final received signal at the $t$-th time frame, the $p$-th subcarriers, and the $q$-th time slots is given by
\begin{equation}
{y_{p,q}}(t) = {\bf{w}}^{\mathsf H}{{\bf{H}}_{{\rm{s,}}p}}{{\bf{\Omega }}_q}{{\bf{H}}_{{\rm{u,}}p}}{{\bf{x}}_p}(t) + {n_{p,q}}(t),
\label{eq_6}
\end{equation}
where ${n_{p,q}}(t)$ is the Gaussian noise. The phase-shift vector in the $q$-th time slot is ${{\bf{v}_q}}$, and its diagonalization can be represented as ${{\bf{\Omega }}_q} = \mathrm{diag}\left( {{\bf{v}_q}} \right)$. Defining ${{\bf{\tilde h}}_p} = {\bf{w}}^{\mathsf H}{{\bf{H}}_{{\rm{s,}}p}}\in \mathbb{C}^{1 \times N_{\rm r}}$, we rewrite Eqn.~\eqref{eq_6} as 
\begin{align}
{y_{p,q}}(t){\kern 1pt}  =& {{\bf{\tilde h}}_p}{{\bf{\Omega }}_q}{{\bf{H}}_{{\rm{u,}}p}}{\bf{x}}_p(t) + {n_{p,q}}(t)\notag \\
= &{{\bf{v}}_q}\mathrm{diag}({{\bf{\tilde h}}_p}){{\bf{H}}_{{\rm{u,}}p}}{{\bf{x}}_p}(t) + {n_{p,q}}(t).
\end{align}

To express ${{\mathbf{H}}_{{\mathrm{u,}}p}}$ more concisely, we define ${\bm {{ \Phi}}}_{q,p}  = {{\mathbf{v}}_q}\mathrm{diag}({{\mathbf{\tilde h}}_p})\in \mathbb{C}^{1 \times N_{\mathrm r}}$ and expand channel ${{\mathbf{H}}_{{\mathrm{u,}}p}}$ to obtain
\begin{align}
 {y_{p,q}}(t){\kern 1pt} = \sum\limits_{l = 1}^{{L}} {{\gamma_l}} {e^{ - j2\pi {f_{\rm{s}}}{\tau_l}\frac{p}{{{P_0}}}}}{\bm {{\Phi}}}_{q,p}{{\mathbf{a}}_{{\rm{IRS}}}}\left( {{\theta _{{\mathrm{e}},l}},{\phi_{{\rm{a}},l}},{u_l}} \right)\notag\\
   {\bf{a}}_{{\rm{ue}}}^{\mathsf H}\left( {{\psi_l}} \right){{\bf{x}}_p}(t) + {n_{p,q}}(t).
\end{align}

After all the time slots $Q$, the received signal at the $t$-th time frame with the $p$-th subcarrier is expressed as
\begin{align}
 {{\bf{y}}_p}(t){\kern 1pt}  = \sum\limits_{l = 1}^{{L}} {{\gamma _l}} {e^{ - j2\pi {f_{\rm{s}}}{\tau _l}\frac{p}{{{P_0}}}}}{{{\bf{\tilde V}}}_p}{{\bf{a}}_{{\rm{IRS}}}}\left( {{\theta _{{\rm{e}},l}},{\phi _{{\rm{a}},l}},{u_l}} \right)\notag\\
  {\bf{a}}_{{\rm{ue}}}^{\mathsf H}\left( {{\psi _l}} \right){{\bf{x}}_p}(t)+ {n_p}(t),
\end{align}

where ${\mathbf{y}_p}\left( t \right) = {\left[ {{y_{p,1}}\left( t \right), \ldots ,{y_{p,Q}}\left( t \right)} \right]^{\mathsf T}}\in \mathbb{C}^{Q \times 1}$, ${{{\bf{\tilde V}}}_p} = {\left[ {\bm{{\Phi}}^{\mathsf T}_{1,p}, \ldots ,\bm{{\Phi}}^{\mathsf T}_{Q,p}} \right]^{\mathsf T}}\in \mathbb{C}^{Q \times N_{\rm r}}$. Thus, we have ${{\bf{x}}_p}(t) = {\bf{x}}(t)$ and ${{{\bf{\tilde V}}}_p} = {\bf{\tilde V}}$. After $T_{{\rm{a}}} $ time frames, the received signal at the $p$-th subcarrier is expressed as
\begin{align}
  {{\bf{Y}}_p} = \sum\limits_{l = 1}^{{L}} {{\gamma _l}} {e^{ - j2\pi {f_{\rm{s}}}{\tau  _l}\frac{p}{{{P_0}}}}}{{\bf{\tilde a}}_{{\rm{IRS}}}}\left( {{\theta _{{\rm{e}},l}},{\phi _{{\rm{a}},l}},{u_l}} \right)\notag\\
   {\bf{\tilde a}}_{{\rm{ue}}}^{\mathsf H}\left( {{\psi _l}} \right) + {{\bf{N}}_p},
    \label{eq_10}
\end{align}
where ${\bf{\tilde a}}_{{\rm{ue}}}\left( {{\psi _l}} \right) = {{\bf{F}}^{\mathsf H}}{\bf{a}}_{{\rm{ue}}}\left( {{\psi _l}} \right)\in \mathbb{C}^{T_{\rm a} \times 1}$, ${\mathbf{F}} = [{\mathbf{x}}(1), \ldots, {\mathbf{x}}(T_{{\mathrm{a}}} )]\in \mathbb{C}^{N_{\mathrm t} \times T_{\mathrm a}}$ collects the beamforming vectors for all time frames, ${{\bf{\tilde a}}_{{\rm{IRS}}}}\left( {{\theta _{{\rm{e}},l}},{\phi _{{\rm{a}},l}},{u_l}} \right) = {{{\bf{\tilde V}}}}{{\bf{a}}_{{\rm{IRS}}}}\left( {{\theta _{{\rm{e}},l}},{\phi _{{\rm{a}},l}},{u_l}} \right)\in \mathbb{C}^{Q \times 1}$, and ${{\bf{N}}_p} = [{n_p}(1), \ldots {n_p}(T_{{\rm{a}}} )]\in \mathbb{C}^{Q \times T_{\rm a}}$.

\begin{figure*}[t]
\centering
\subfloat[Harmony composition-inspired NF IRS channel parameter estimation framework.]{%
    \label{subfig_3a} 
    \includegraphics[width=6in]{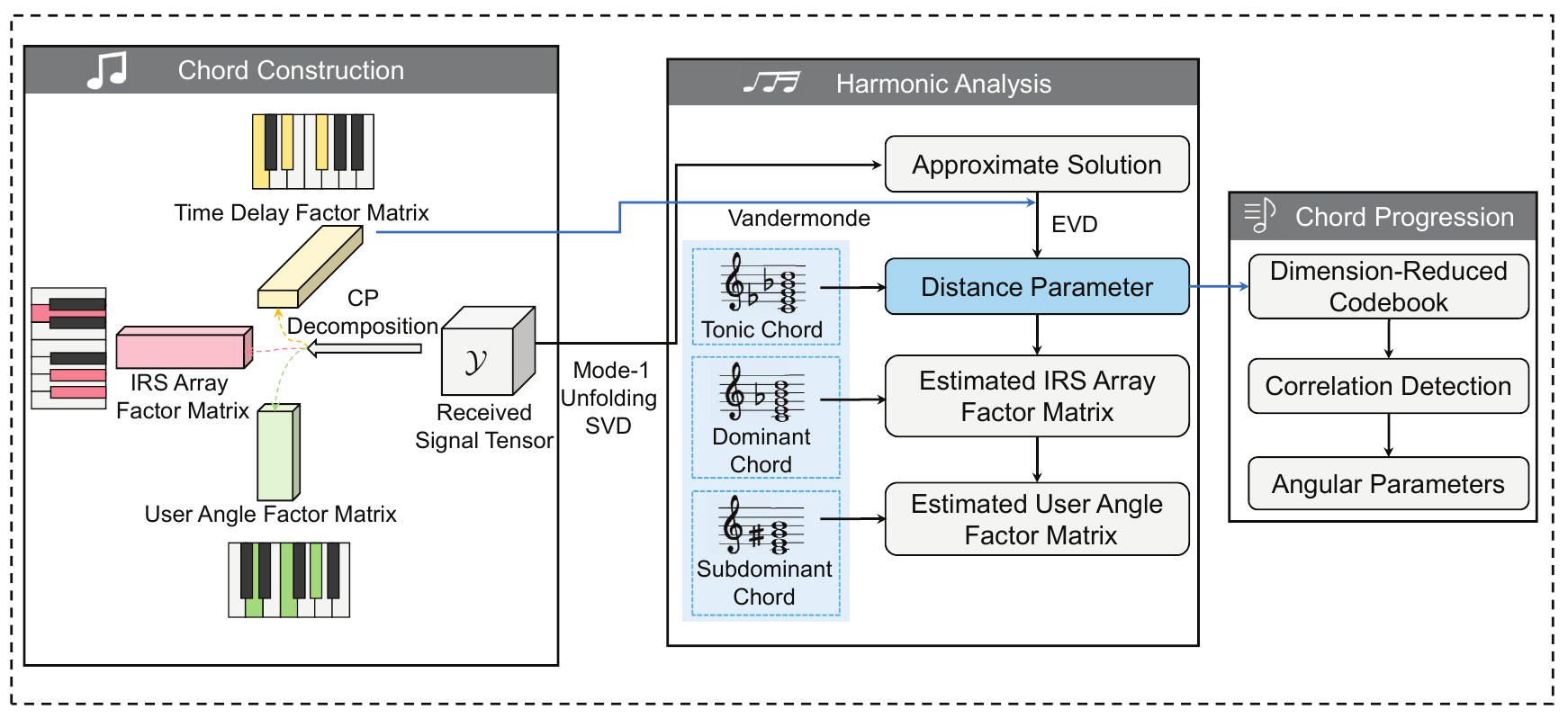}%
}\\ 
\vspace{0.3cm} 
\subfloat[Mapping of physical parameters to CP tensor modes.]{%
    \label{subfig_3b} 
    \includegraphics[width=6in]{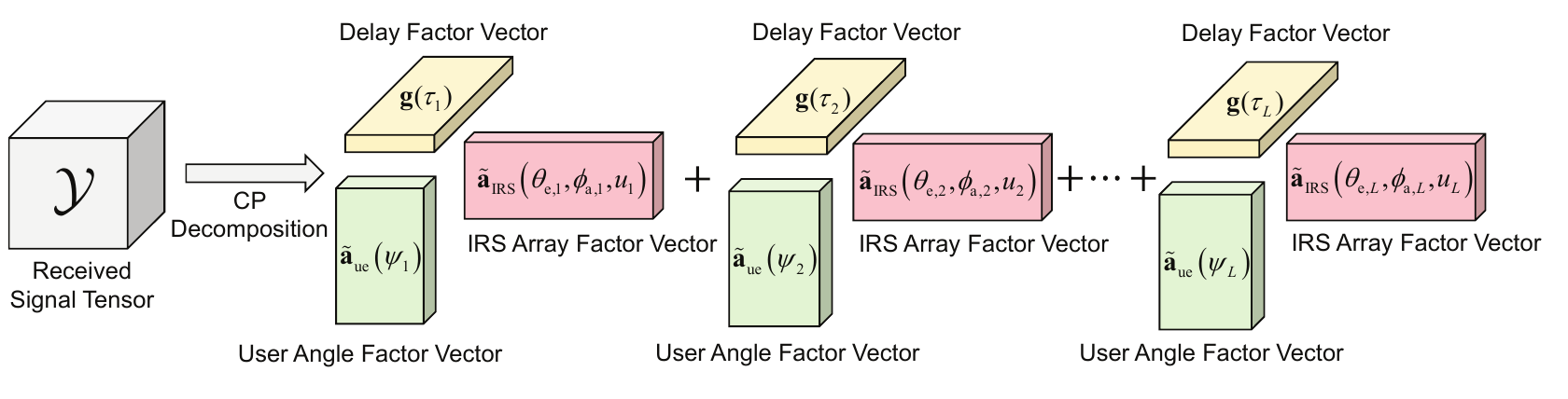}%
}
\caption{Harmony-based NF IRS channel estimation framework and underlying physical parameters mapping to CP tensor modes}
\label{fig_3}
\end{figure*}

\section {Harmonic Processing-Inspired NF IRS Channel Parameter Estimation Framework} \label{III}
This section proposes a harmonic processing-inspired channel estimation framework for NF IRS-assisted communications. As shown in Fig.~\ref{fig_3}(a), the tensor
modalization-based framework mainly consists of three steps: chord construction, harmonic analysis, and chord progression. 
To be specific, the time-delay term, NF IRS array response, and user angle array response are constructed into three different factor matrices, as realized in the chord construction step. The estimation of the distance parameters is performed in the harmonic analysis step. The extraction of the AoA/AoD parameters is realized in the chord progression step.
In what follows, we will provide a detailed description of these steps for channel estimation.

\subsection{Chord Construction}
During the harmonization process, constructing different chords upon a shared note creates smoother harmonic progressions. Similarly, in our approach, we observe that both the delay term and the IRS array response structures share the same distance parameter, as demonstrated in \eqref{eq_10}. By strategically assigning this shared feature to different tensor dimensions, we can effectively decouple the distance parameter through tensor CP decomposition. 

Based on \eqref{eq_10}, the signal received from all subcarriers can be expressed as a third-order tensor ${{\bf{\cal Y}}}\in {\mathbb{C}^{Q \times T_{{\rm{a}}}   \times P}}$, which is written as
\begin{equation}
{{\bf{\cal Y}} = \mathop \sum \limits_{l = 1}^{{L}} {{\bf{\tilde a}}_{{\rm{IRS}}}}\left( {{\theta _{{\rm{e}},l}},{\phi _{{\rm{a}},l}},{u_l}} \right) \circ\left( {{\gamma _l}{{{\bf{\tilde a}}}_{{\rm{ue}}}}\left( {{\psi _l}} \right)} \right)  \circ  {{\bf{g}}\left( {{\tau_l}} \right)}  + {\bf{\cal N}}},
\label{eq_11}
\end{equation}
where ${{\bf{\cal N}}}\in {\mathbb{C}^{Q \times T_{{\rm{a}}}  \times P}}$ denotes the tensor noise, and ${\bf{g}}({\tau_l}) = {\left[ {{e^{ - j2\pi \frac{{{f_{\rm{s}}}}}{{{P_0}}}{\tau_l}}}, \ldots ,{e^{ - j2\pi \frac{{{f_{\rm{s}}}}}{{{P_0}}}P{\tau _l}}}} \right]^{\mathsf T}}\in \mathbb{C}^{P \times 1}$. { In fact, the tensor rank $L$ is typically treated as prior knowledge in the channel estimation framework, as shown in~\cite{b2}. Of course, the tensor rank $L$ can also be estimated using the minimum description length (MDL) criterion, as described in ~\cite{b3}.} 
Based on the constructed form of the reception tensor ${\bf{\cal Y}}$, we can write its factor matrix
\begin{align}
{\bf{A}} &= \left[ {{\bf{\tilde a}}_{{\rm{IRS}}}\left( {{\theta _{{\rm{e}},1}},{\phi _{{\rm{a}},1}},{u_1}} \right), \ldots ,{\bf{\tilde a}}_{{\rm{IRS}}}\left( {{\theta _{{\rm{e}},L}},{\phi _{{\rm{a}},L}},{u_L}} \right)} \right]\in \mathbb{C}^{Q \times L},\notag\\[5pt]
{\bf{B}} &= \left[ {{\gamma _1}{{{\bf{\tilde a}}}_{{\rm{ue}}}}\left( {{\psi _1}} \right), \ldots ,{{\gamma _{{L}}}{{\bf{\tilde a}}}_{{\rm{ue}}}}\left( {{\psi _L}} \right)} \right]\in \mathbb{C}^{T_{\rm{a}} \times L},\notag\\[5pt]
{\bf{C}}& = \left[ {{\bf{g}}\left( {{\tau_1}} \right), \ldots ,{\bf{g}}\left( {{\tau_{{L}}}} \right)} \right]\in \mathbb{C}^{P \times L}, 
\label{eq_12}
\end{align}
where $\mathbf{A}$, $\mathbf{B}$, and $\mathbf{C}$ denote the IRS array factor matrix, the user angle factor matrix, and the time-delay factor matrix, respectively.
{ The proposed method draws on the principle in music theory of breaking down complex chords into independent intervals ~\cite{M1,MM2}. In harmonic construction, the tonic chord serves as the tonal center, representing stability and foundation. This corresponds to the delay factor matrix, which contains the most critical distance parameters in the NF channel, and is therefore classified as the \emph{tonic chord}. The dominant chord, by contrast, is filled with dissonance and tension. Similarly, the IRS array factor matrix encompasses both distance and angle parameters, greatly increasing system complexity and dimensionality, thereby creating rich ``variations'' and ``possibilities''. This multidimensional parameter coupling, much like the tension of the dominant chord, strongly drives changes and evolution in the system state, and is thus analogized to the \emph{dominant chord}. The subdominant chord, serving as a transitional element in harmony and imparting smoothness or turning points, is comparable to the user angle factor matrix in a NF IRS system, which plays a similar auxiliary role. Accordingly, we classify it as the \emph {subdominant chord}.
} 

{ In the CP tensor decomposition structure, a factor matrix representing a specific feature information. Each factor matrix is composed of several factor vectors, with each vector corresponding to a propagation path. The factor vector under each propagation path characterizes the specific parameter features of that path. For the delay-related factor vector ${\bf{g}}({\tau_l}) = {\left[ {{e^{ - j2\pi \frac{{{f_{\rm{s}}}}}{{{P_0}}}{\tau_l}}}, \ldots ,{e^{ - j2\pi \frac{{{f_{\rm{s}}}}}{{{P_0}}}P{\tau _l}}}} \right]^{\mathsf T}}\in \mathbb{C}^{P \times 1}$, it can be observed that the delay term ${e^{ - j2\pi {f_{\rm{s}}}{\tau  _l}\frac{p}{{{P_0}}}}}$ is a function of the subcarrier index $p$. Therefore, the factor vector associated with the delay is mainly mapped to the frequency dimension. For the IRS array factor vector ${{\bf{\tilde a}}_{{\rm{IRS}}}}\left( {{\theta _{{\rm{e}},l}},{\phi _{{\rm{a}},l}},{u_l}} \right) = {{{\bf{\tilde V}}}}{{\bf{a}}_{{\rm{IRS}}}}\left( {{\theta _{{\rm{e}},l}},{\phi _{{\rm{a}},l}},{u_l}} \right)\in \mathbb{C}^{Q \times 1}$ that contains AoA features, while ${{{\bf{\tilde V}}}}$ represents a matrix that captures the effects of all time slots, i.e., its row dimension corresponds to the number of time slots. When ${{{\bf{\tilde V}}}}$ is combined with the array response ${{\bf{a}}_{{\rm{IRS}}}}\left( {{\theta _{{\rm{e}},l}},{\phi _{{\rm{a}},l}},{u_l}} \right)$, it forms a time-slot-dependent IRS angular response factor vector.
Similarly, the factor vector ${\bf{\tilde a}}_{{\rm{ue}}}\left( {{\psi _l}} \right) = {{\bf{F}}^{\mathsf H}}{\bf{a}}_{{\rm{ue}}}\left( {{\psi _l}} \right)\in \mathbb{C}^{T_{\rm a} \times 1}$ containing AoD features is constructed by combining the user-side array response ${\bf{a}}_{{\rm{ue}}}\left( {{\psi _l}} \right)$ with the matrix $ {\mathbf{F}}$, which accounts for the effects of all time frames, i.e., its column dimension depends on the number of time frames. The characteristic basis vectors associated with different physical parameters form a rank-1 tensor for each propagation path via outer products. The superposition of all rank-1 tensors across all paths constitutes the CP structure of the observed signal tensor, as illustrated in Fig.~\ref{fig_3}(b). In this CP structure, the outer-product operation allows each factor vector to control only the variation within its corresponding mode, thereby ensuring that the influence of different physical parameters is mainly concentrated in their respective modes.}

\subsection{Harmonic Analysis}
 {In harmonic analysis, the tonic provides stability, the dominant generates tension \footnote{{In music theory, the tension inherent in dominant chords manifests as harmonic instability and a compelling drive toward resolution. This tension stems from an internal entanglement of tonal components that is difficult to isolate, a phenomenon that parallels the mathematical characteristics of near-field channels, where distance and angular parameters are intrinsically coupled. This intrinsic state of intertwined dependencies, which necessitates untangling, motivates the subsequent decoupling mechanisms in near-field signal processing.
}}, and the subdominant serves as a transition. The functional interplay among these three chords evokes specific emotional experiences and enhances musical expressiveness. This process is reflected in our channel estimation method. By performing SVD on the mode-1 unfolding of the observed signal tensor and exploiting the inherent Vandermonde structure of the delay factor matrix, we transform the SVD result into an EVD to accurately estimate the distance parameters. The resulting delay factor matrix (tonic chord) effectively supports the estimation of the IRS array factor matrix (dominant chord) and the user angle factor matrix (subdominant chord). Inspired by the leading role of the tonic in harmonic progression, we construct a distance-related yet dimension-reduced codebook based on the estimated delay factor matrix. Through correlation detection with the estimated IRS array factor matrix or user angle factor matrix, AoAs and AoDs can be efficiently extracted. This efficient parameter estimation process, analogous to harmonic expression guided by the tonic, achieves more accurate channel parameter estimation. Drawing inspiration from harmonic analysis, our method transforms the complex, high-dimensional NF IRS channel estimation problem into a low-dimensional and independent factor matrix optimization problem, thereby enabling closed-form solutions for the channel parameter factor matrices. Unlike baseline algorithms, our approach eliminates the need for iterative high-dimensional codebook searches when extracting angular parameters. By decoupling the distance parameters and utilizing low-dimensional factor matrices, our method supports higher-resolution codebooks and significantly improves estimation accuracy.
} 
\begin{figure}[t]
\centering
\includegraphics[width=2.9in]{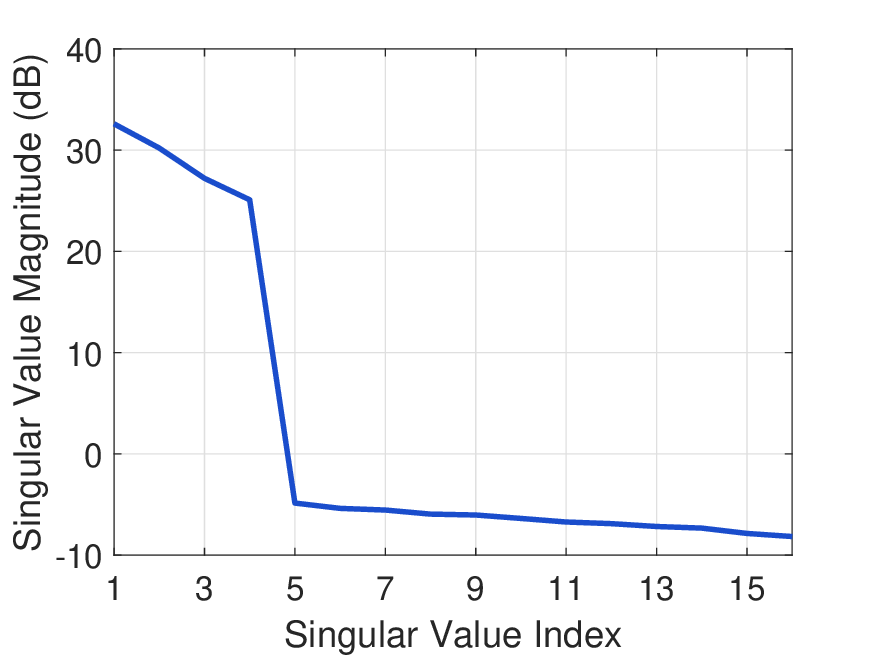}
\caption{{The singular value distribution characteristics.}}
\label{fig_s}                                                  
\end{figure}
Initially, we need to determine the tonic chord, which corresponds to the distance parameter.
According to~\cite{r23, r24}, tensor CP decomposition is unique when one of the tensor's factor matrices has a Vandermonde structure and the individual factor matrices' dimensions meet the condition of $\min \left( {\left( {P - 1} \right)T_{{\rm{a}}} ,Q} \right) \ge {L}$. Then, we perform model-1 expansion~\cite{r25} of the tensor ${\bf{\cal Y}}$, as given by
\begin{equation}
{\bf{Y}}_{(1)}^{\mathsf T} = \left( {{\bf{C}} \odot {\bf{B}}} \right){{\bf{A}}^{\mathsf T}} + {\bf{N}}_{(1)}^{\mathsf T},
 \label{eq_13}
\end{equation}
where ${\bf{N}}_{(1)}\in \mathbb{C}^{Q \times T_{\rm a}P}$ represents the model-1 expansion form of the noise tensor ${{\bf{\cal N}}}$.

Applying SVD to Eqn.~\ref{eq_13} yields
\begin{equation}
{\bf{Y}}_{(1)}^{\mathsf T} = {\bf{U\Sigma }}{{\bf{V}}^{\mathsf H}},
 \label{eq_14}
\end{equation}
where ${\bf{U}} \in {\mathbb{C}^{T_{{\rm{a}}} P \times {L}}}$ and ${\bf{V}} \in {\mathbb{C}^{Q \times {L}}}$ are eigenvectors and ${\bf{\Sigma }} \in {\mathbb{C}^{{L} \times {L}}}$ denotes the singular value. { Since the sum of the rank-one tensors of all paths forms the CP decomposition of the entire received signal tensor, the singular value distribution exhibits the characteristic where the first $L$ singular values are large, while the remaining singular values are small and decay slowly, as shown in Fig.~\ref{fig_s}.}
 Neglecting the effect of noise, it can be deduced from \eqref{eq_13} and \eqref{eq_14} that there exists a nonsingular matrix ${\bf{M}}\in \mathbb{C}^{L \times L}$ such that 
\begin{equation}
{\bf{UM}} = {\bf{C}} \odot {\bf{B}}.
 \label{eq_15}
\end{equation}

{ Due to the fixed delay, the phase increases uniformly and linearly with the subcarrier index, so the delay factor matrix ${\bf{C}}$ has a Vandermonde structure
}, we have
\begin{equation}
\left( {{\underline{\bf{C}}} \odot {\bf{B}}} \right){\bf{Z}} = {\overline{\bf{C}}} \odot {\bf{B}},
\label{eq_16}
\end{equation}
where ${\underline{\bf{C}}}$  is a submatrix of ${\bf{C}}$ ,and its form is obtained by removing the  the $P$-th row of ${\bf{C}}$, ${\overline{\bf{C}}}$ is obtained by removing the first row of ${\bf{C}}$.
${\bf{Z}} = {\mathrm{diag}}({z_1}, \ldots ,{z_{{L}}})\in {\mathbb{C}^{L \times {L}}}$ and ${z_l} = {e^{ - j2\pi \frac{{{f_{\rm{s}}}}}{{{P_0}}}{\tau_l}}}$, ${z_l}$ is a constituent of the factor matrix ${\bf{C}}$. Recalling~\eqref{eq_15}, we have
\begin{align}
{{\bf{U}}_1}{\bf{M}} = {{\underline{\bf{C}}}} \odot {\bf{B}},\notag\\
{{\bf{U}}_2}{\bf{M}} = {{\overline{\bf{C}}}} \odot {\bf{B}},
\label{eq_17}
\end{align}
where ${{\bf{U}}_1}$ and ${{\bf{U}}_2}$ are expressed as
\begin{align}
{{\bf{U}}_1} &= {\bf{U}}\left( {1:\left( {P - 1} \right)T_{{\rm{a}}} ,:} \right)\in {\mathbb{C}^{T_{{\rm{a}}} (P-1) \times {L}}},\notag\\  
{{\bf{U}}_2} &= {\bf{U}}\left( {T_{{\rm{a}}}  + 1:PT_{{\rm{a}}} ,:} \right)\in {\mathbb{C}^{T_{{\rm{a}}} (P-1) \times {L}}},  
\end{align}
Combining \eqref{eq_16} and \eqref{eq_17}, we obtain
\begin{equation}
{{\bf{U}}_2}{\bf{M}} = {{\bf{U}}_1}{\bf{MZ}}.
\label{eq_19}
\end{equation}

Based on the SVD property, it can be concluded that 
${{\bf{U}}_1}$ and ${{\bf{U}}_2}$ are column full rank. Eqn.~\eqref{eq_19} can be written in the EVD form as
\begin{equation}
{\bf{U}}_1^\dag {{\bf{U}}_2} = {\bf{\hat M\hat Z}}{{{\bf{\hat M}}}^{ - 1}}. 
\label{eq_20}
\end{equation}

From \eqref{eq_20}, we can estimate ${\bf{Z}}$. Since the elements of ${\bf{Z}}$ are also contained in the matrix ${\hat{\bf{C}}}$, we obtain
\begin{equation}
{\bf{{\hat c}}}_l = \left[ {{{\hat z}_l^1},{\kern 1pt} {\kern 1pt} {\kern 1pt} \hat z{\kern 1pt} _l^2{\kern 1pt} , \ldots ,\hat z{\kern 1pt} _l^P} \right]^{\mathsf T}. 
\label{eq_21}
\end{equation}
According to \eqref{eq_15}, the estimated elements of the factor matrix ${\hat{\bf{B}}}$ can be obtained as
\begin{equation}
{\bf{{\hat b}}}_l = \left( {\frac{{\bf{{{\hat c}}}}^{\mathsf H}_l}{{{\bf{{{\hat c}}}}^{\mathsf H}_l{\bf{{\hat c}}}}_l} \otimes {{\bf{I}}_{\rm{T}}}} \right){\bf{U\hat M}}\left( {:,l} \right).
\label{eq_22}
\end{equation}
Given the estimated factor matrices ${\hat{\bf{B}}}$ and ${\hat{\bf{C}}}$, we use \eqref{eq_13} and obtain
\begin{equation}
{\bf{\hat A}} = {{\bf{Y}}_{(1)}}{\left( {{{\left( {{\bf{\hat C}} \odot {\bf{\hat B}}} \right)}^{\mathsf T}}} \right)^\dag }.
\label{eq_23}
\end{equation}

\subsection{Chord Progression}
In music, chord progressions are structured around the tonic chord to establish harmonic stability and maintain rhythmic continuity. Similarly, we propose a channel estimation technique in which an estimated time-delay factor matrix serves a foundational role, akin to the tonic chord.
We first design a dimension-reduced NF channel array response codebook based on the time-delay factor matrix.
{Subsequently, we evaluate the correlation between the estimated IRS array or user angle factor matrix and our specifically designed codebook to extract the channel angle parameters.}

Based on \eqref{eq_21}, the time delay parameter ${{\hat \tau}_l}$ is rewritten as
\begin{equation}
{{\hat \tau}_l} = - \frac{{{P_0}}}{{2\pi {f_{\rm{s}}}}}\arg \left( {{{\hat z}_l}} \right),
\label{eq_24}
\end{equation}
where $\arg \left({{{\hat z}_l}} \right)$ denotes the argument of ${{\hat z}_l}$. Based on the relationship between delay and distance, the estimated distance is expressed by ${\hat{u}_l}={\hat{\tau}_l}c$. { The dimensionality-reduced NF channel codebook is derived from the full spherical codebook, which covers the the parameters of  distance, azimuth and elevation angles. Since the distance parameter has been estimated using the Vandermonde structure of the delay factor matrix, the constructed codebook now only contains azimuth and elevation angles, resulting in a reduced-dimensional NF channel codebook}. This reduced codebook $\Omega_{R}$ can then be used to extract the angular parameters from the NF IRS array response, as given by

\begin{equation}
{{\Omega}}_{\rm {R}} = \left\{ ({{\theta _{{\rm{c}}}}}, {{\phi _{{\rm{c}}}}}) \,\middle|\, 
\begin{aligned}
&{{\theta _{{\rm{c}}}}} = {{\theta _{{\rm{c}},{\text{min}}}}} + i \Delta{{\theta _{{\rm{c}}}}},  \ i = 1, 2, \ldots, G_{\rm{z}}, \\
&{{\phi _{{\rm{c}}}}} =  {{\phi _{{\rm{c}},{\text{min}}}}} + j \Delta{{\phi _{{\rm{c}}}}},  \ j = 1, 2, \ldots, G_{\rm{y}}
\end{aligned} 
\right\},
\label{eq_25}
\end{equation}
where ${{G_{\rm{z}}}}$ and ${{G_{\rm{y}}}}$ represent the dimensions of the two angles in the angular codebook, ${{\theta _{{\rm{c}},{\text{min}}}}}({{\phi _{{\rm{c}},{\text{min}}}}})$  denotes the minimum elevation (azimuth) values, and $\Delta{{\theta _{{\rm{c}}}}}(\Delta{{\phi _{{\rm{c}}}}})$ indicates the sampling step sizes for elevation (azimuth).
{ The sampling step size directly determines the angular resolution of the codebook. The smaller the step size, the finer the division of the codebook in terms of angle, resulting in higher resolution. Conversely, the larger the step size, the lower the resolution. A higher codebook resolution can significantly improve the accuracy of angle estimation. The principle behind this is that a denser codebook can more accurately approximate the true angle values, thereby effectively reducing quantization errors and enhancing estimation performance.
} 
The estimators of the elevation and azimuth angles can then be expressed as
\begin{equation}
{\kern 1pt} {\kern 1pt} \{{\hat \theta }_{{\rm{r}},l},{\hat \phi }_{{\rm{r}},l}\} = \arg \mathop {\max }\limits_{{\theta _{{\rm{c}}}},{\phi _{{\rm{c}}}}\in\Omega_{R}}  {\kern 1pt} \frac{{\left| {{\bf{\hat a}}_{{\rm{IRS}},l}^{\mathsf H}{{{\bf{\tilde a}}}_{\rm{IRS}}}\left( {{\theta _{{\rm{c}}}},{\phi _{{\rm{c}}}},{\hat{\mu} _l}} \right)} \right|^2 }}{{{{\left\| {{{{\bf{\hat a}}}_{{\rm{IRS}},l}}} \right\|}_2}{{\left\| {{{{\bf{\tilde a}}}_{\rm{IRS}}}\left( {{\theta _{{\rm{c}}}},{\phi _{{\rm{c}}}},{\hat{\mu} _l}} \right)} \right\|}_2}}},
\label{eq_26}
\end{equation}
where ${{{{\bf{\hat a}}}_{{\rm{IRS}},l}}}$ is the $l$-th column of ${{\bf{\hat A}}}$. Similar to~\eqref{eq_25}, the user angle codebook set is given by $\Omega_{{\rm{U}}} = \left\{ {{\psi _{{\rm{c}}}}}  \,\middle|\, 
\begin{aligned}
&{{\psi _{{\rm{c}}}}} = {{\psi _{\rm{c,}{\text{min}}}}} + k\Delta{{\psi _{{\rm{c}}}}},  \ k = 1, 2, \ldots, G_{\rm{u}}, \\
\end{aligned} 
\right\}$. 
${{\psi _{\rm{c,}{\text{min}}}}}$, $\Delta{{\psi _{{\rm{c}}}}}$ and $G_{\rm{u}}$ represent the minimum angle values in the codebook, angular sampling step size, and angular codebook dimension, respectively. The estimator of the angle ${{\hat \psi_l}}$ is given by
\begin{equation}
{{\hat \psi }_l} = \arg \mathop {\max }\limits_{{\psi_{\rm{c}} }\in\Omega_{{\rm{U}}} } {\kern 1pt} \frac{{\left| {{\bf{\hat b}}_l^{\mathsf H}{{{\bf{\tilde a}}}_{{\rm{ue}}}}\left( {{{{\psi _{{\rm{c}}}}}}} \right)} \right|^2}}{{{{\left\| {{{{\bf{\hat b}}}_l}} \right\|}_2}{{\left\| {{{{\bf{\tilde a}}}_{{\rm{ue}}}}\left( {{{{\psi _{{\rm{c}}}}} }} \right)} \right\|}_2}}},
\label{eq_27}
\end{equation}
where ${{{{\bf{\hat b}}}_l}}$ represents the columns of the factor matrix ${\bf{B}}$.

Finally, we estimate the path gain ${\gamma_l}$ based on the known parameters. 
The factor matrices ${{{\bf{\bar A}}}_{{\rm{IRS}}}} = \left[ {{{{\bf{\tilde a}}}_{\rm{IRS}}}\left( {{{\hat \theta }_{{\rm{e}},1}},{{\hat \phi }_{{\rm{a}},1}},{{\hat \mu }_1}} \right), \ldots ,{{{\bf{\tilde a}}}_{\rm{IRS}}}\left( {{{\hat \theta }_{{\rm{e}},L}},{{\hat \phi }_{{\rm{a}},L}},{{\hat \mu }_L}} \right)} \right]$, ${{{\bf{\bar B}}}_{{\rm{ue}}}} = \left[ {{{{\bf{\tilde a}}}_{{\rm{ue}}}}\left( {{\psi _1}} \right), \ldots ,{{{\bf{\tilde a}}}_{{\rm{ue}}}}\left( {{\psi _L}} \right)} \right]$, and ${\bf{\bar C}} = \left[ {{\bf{g}}({{\hat \tau}_1}), \ldots ,{\bf{g}}({{\hat \tau}_{{L}}})} \right]$ are created by combining the obtained parameters. The model-2 expansion of the received signal tensor is given by
\begin{equation}
{\bf{Y}}_{(2)}^{\mathsf T} = \left( {{\bf{C}} \odot {\bf{A}}} \right){{\bf{B}}^{\mathsf T}} + {\bf{N}}_{(2)}^{\mathsf T}.
\end{equation}

Neglecting the noise, the estimated factor matrix is written by ${\hat{\bf{ B}}} = {{\bf{Y}}_{(2)}}{\left( {{{\left( {{{{\bf{\bar C}}}} \odot {{{\bf{\bar A}}}_{{\rm{IRS}}}}} \right)}^{\mathsf T}}} \right)^\dag }$. According to \eqref{eq_12}, we can obtain the estimate of the path gain as follows:
\begin{equation}
{\hat{\bm{\upgamma }}} = {{{{\bf{\bar B}}}_{{\rm{ue}}}}^\dag }{\hat{\bf{ B}}},
\label{eq_29}
\end{equation}
where ${\hat{\bm{\upgamma }}} = \mathrm{diag}\left( {{\gamma _1}, \ldots ,{\gamma _{{L_u}}}} \right)$. After the above operations, we complete the channel parameter estimation. The detailed algorithm is presented in Algorithm~\ref{alg:alg1}.

\noindent{{\textbf{Remark:} Regarding the potential degradation of the estimation performance of our scheme under conditions such as noise, quantization, or hardware defects, first, for high noise environments, we can use a feature-value weighted robust SVD algorithm for denoising ~\cite{rk1}. Second, for quantization distortion, employing oversampling is a feasible solution to mitigate the impact of quantization distortion~\cite{rk2}. For degradation caused by systematic distortions introduced by hardware defects, we can compensate by establishing nonlinear models for components such as power amplifiers~\cite{rk3}. Finally, for the non-uniqueness of CPD caused by coherent paths or rank-deficient channels, we can address the rank deficiency issue through a pre-processing spatial smoothing module~\cite{rk4}.} }

\begin{algorithm}[htbp]
\caption{Harmonic Processing-Inspired NF IRS Channel Estimation}
\label{alg:alg1}
\begin{algorithmic}
\STATE 
\REQUIRE Received signal ${{\bf{Y}}_p}$.
\\
\textbf{Step 1: Chord Construction
}\\
1: Construct a tensor of receive signals, ${{\bf{\cal Y}}}$. \\
2: Design factor matrices (chords) ${\bf{A}}$, ${\bf{B}}$, and ${\bf{C}}$ from ${{\bf{\cal Y}}}$.\\
\textbf{Step 2:  Harmonic Analysis}\\
3: Tensor mode-1 unfold   ${\bf{Y}}_{(1)}^{\mathsf T} = \left({{\bf{C}}\odot {\bf{B}}}\right){{\bf{A}}^{\mathsf T}} + {\bf{N}}_{(1)}^{\mathsf T}$. \\
4: SVD of unfolded vectors   ${\bf{Y}}_{(1)}^{\mathsf T} = {\bf{U\Sigma }}{{\bf{V}}^{\mathsf H}}$.\\
5: Construct the submatrix of ${\bf{U}}$:\\
\STATE \hspace{0.5cm}
${{\bf{U}}_1} ={\bf{U}}\left( {1:\left( {P - 1} \right)T_{{\rm{a}}} ,:} \right)$,\\
\STATE \hspace{0.5cm}
${{\bf{U}}_2} = {\bf{U}}\left( {T + 1:PT_{{\rm{a}}} ,:} \right)$. \\
6: Calculate the EVD: ${\bf{U}}_1^\dag {{\bf{U}}_2} = {\bf{\hat M\hat Z}}{{{\bf{\hat M}}}^{-1}}$. \\
7: Estimate ${{\hat z}_l} \to {\bf{\hat C}}$ (\textbf {tonic chord}).\\    
8: Obtain
$\left\{ {{{\bf{{\hat b}}}_l}} \right\}_{l = 1}^{{L}} \to {\bf{\hat B}}$ (\textbf {subdominant chord}) using \eqref{eq_22}.\\
9: Obtain  ${\bf{\hat A}}$ (\textbf {dominant chord}) using\eqref{eq_23}.\\
\textbf{Step 3: Chord Progression}\\
10: Establish the codebook ${{\Omega}}_{\rm{R}}$ and ${{\Omega}}_{\rm{U}}$.\\
Estimate $ {{{\hat \tau}_l}}$, ${{\hat \theta _{{\rm{e}},l}}}$, ${\hat \phi _{{\rm{a}},1}}$, ${{\hat \psi _l}}$ using \eqref{eq_24}, \eqref{eq_26},\eqref{eq_27}. \\
11: Estimate ${{{\hat \gamma }_l}}$ using \eqref{eq_29}.  \\
12: \textbf{return}
Channel parameters $\left\{ {{\hat\theta _{{\rm{e}},l}},{\hat \phi _{{\rm{a}},l}},{\hat \tau_l},{{\hat \psi _l}},{\hat \gamma _l}} \right\}_{l = 1}^{{L}}$.
\end{algorithmic}
\label{alg1}
\end{algorithm}

\section{Performance Analytics}\label{IV}
\subsection{Cram\'{e}r-Rao Lower Bound}
Since the CRLB serves as a common bound for all unbiased estimates~\cite{r27, r28}, this section derives CRLB based on~\eqref{eq_11} to assess the proposed approach.

The parameter set that needs to be estimated is referred to as $\left\{ {{\hat\theta _{{\rm{e}},l}},{\hat \phi _{{\rm{a}},l}},{\hat \tau_l},{{\hat \psi _l}},{\hat \gamma _l}} \right\}_{l = 1}^{{L}}$. Assume that each element of ${\mathbf{\mathcal N}}$ is independently and identically distributed (i.i.d.) and takes the value of a symmetric circular Gaussian random variable with mean $0$ and variance ${\sigma ^2}$. The CRLB is derived by computing derivatives for each element, which can be integrated for a more streamlined presentation, such as: ${{\bm{\uptheta}}_{\rm{e}}} = {\left[ {{\theta _{{\rm{e}},1}} ,\ldots, {\theta _{{\rm{e}},L}}} \right]^{\mathsf T}}$, ${{\bm{\upphi}}_{\rm{a}}} = {\left[ {{ \phi _{{\rm{a}},1}}, \ldots, { \phi _{{\rm{a}},L}}} \right]^{\mathsf T}}$, ${{\bm{\uptheta }}}_{\rm{ue}} = {\left[ {{{ \psi _1}} ,\ldots, {{ \psi _L}}} \right]^{\mathsf T}}$, ${\bf{u}} = {\left[ {{\tau_1},\ldots, {\tau_{{L}}}} \right]^{\mathsf T}}$, ${\bm{\upgamma }} = {\left[ {{\gamma _1}, \ldots ,{\gamma _{{L}}}} \right]^{\mathsf T}}$ and ${{\bf{p}}_{\rm{e}}} = \left[ {{\boldsymbol{\uptheta }}_{\rm{e}}^{\mathsf T}, {\boldsymbol{\upphi}}_{\rm{a}}^{\mathsf T}, {\boldsymbol{\uptheta }}_{\rm{ue}}^{\mathsf T}, {{\bf{u}}^{\mathsf T}}, {{\bm{\upgamma }}^{\mathsf T}} } \right]$.

The log-likelihood function for ${{\boldsymbol{p}}_{\rm{e}}}$ is denoted as
\begin{align}
L_{{\rm{o}}} \left( {{\bf{p}}_{\rm{e}}} \right) \overset{\text{a}}{=} & - QT_{{\rm{a}}}P\ln \left( {\pi {\sigma ^2}} \right) - \frac{1}{{{\sigma ^2}}}\left\| {{\bf{Y}}_{(1)}^{\mathsf T} - \left( {{\bf{C}} \odot {\bf{B}}} \right){{\bf{A}}^{\mathsf T}}} \right\|_F^2\notag\\[5pt]
   \overset{\text{b}}{=}&  - QT_{{\rm{a}}}P\ln \left( {\pi {\sigma ^2}} \right) - \frac{1}{{{\sigma ^2}}}\left\| {{\bf{Y}}_{(2)}^{\mathsf T} - \left( {{\bf{C}} \odot {\bf{{A}}}} \right){{\bf{B}}^{\mathsf T}}} \right\|_F^2\notag\\[5pt]
    \overset{\text{c}}{=}&  - QT_{{\rm{a}}}P\ln \left( {\pi {\sigma ^2}} \right) - \frac{1}{{{\sigma ^2}}}\left\| {{\bf{Y}}_{(3)}^{\mathsf T} - \left( {{\bf{B}} \odot {\bf{A}}} \right){{\bf{C}}^{\mathsf T}}} \right\|_F^2,
    \label{eq_30}
\end{align}
where $a$, $b$ and $c$ on equality denote the log-likelihood functions obtained from the model-1, model-2, and model-3 expansions of the tensor $\mathcal{Y}$, respectively. 

To calculate the CRLB, we first need to compute the Fisher information matrix (FIM)~\cite{r30} for the parameter ${{\bf{p}}_{\rm{e}}}$, as given by
\begin{equation}
{\bf{J}}\left( {{\bf{p}}_{\rm{e}}} \right) = \mathbb{E}\left\{ {{{\left( {\frac{{\partial L_{{\rm{o}}}\left( {{\bf{p}}_{\rm{e}}} \right)}}{{\partial {{\bf{p}}_{\rm{e}}}}}} \right)}^{\mathsf H}}\left( {\frac{{\partial L_{{\rm{o}}}\left( {{\bf{p}}_{\rm{e}}} \right)}}{{\partial {{\bf{p}}_{\rm{e}}}}}} \right)} \right\}.
\label{eq_31}
\end{equation}

According to~\eqref{eq_30}, calculating FIM requires the partial derivatives of $L_{{\rm{o}}}\left( {{\bf{p}}_{\rm{e}}} \right)$ with respect to ${{\bf{p}}_{\rm{e}}}$. The partial derivatives with respect to $\left\{ {{\theta _{{\rm{e}},l}},{ \phi _{{\rm{a}},l}},{ \tau_l},{{\psi _l}},{\gamma _l}} \right\}$ are expressed as $\frac{{\partial L_{{\rm{o}}}\left( {{\bf{p}}_{\rm{e}}} \right)}}{{\partial {\theta _{{\rm{e}},l}}}}$,$\frac{{\partial L_{{\rm{o}}}\left( {{\bf{p}}_{\rm{e}}} \right)}}{{\partial { \phi _{{\rm{a}},l}}}}$,$\frac{{\partial L_{{\rm{o}}}\left( {{\bf{p}}_{\rm{e}}} \right)}}{{\partial {{\psi _l}}}}$,$\frac{{\partial L_{{\rm{o}}}\left( {{\bf{p}}_{\rm{e}}} \right)}}{{\partial { \tau_l}}}$, and $\frac{{\partial L_{{\rm{o}}}\left( {{\bf{p}}_{\rm{e}}} \right)}}{{\partial {\gamma _l}}}$, respectively. The detailed derivation of all involved parameters is provided in Appendix~\ref{Appendix A}.

Next, we construct the submatrix of the main diagonal in the FIM, where the $\left(l_1,l_2 \right)$ element of the submatrix corresponding to ${{\bm{\uptheta }}_{\rm{e}}}$ is given by
\begin{align}
{{\bf{J}}_{11}}\left( {{l_1},{l_2}} \right) &= \mathbb{E}\left\{ {{{\left( {\frac{{\partial L_{{\rm{o}}}\left( {{\boldsymbol{p}}_{\rm{e}}} \right)}}{{\partial {\theta _{{\rm{e}},{l_1}}}}}} \right)}^*}\left( {\frac{{\partial L_{{\rm{o}}}\left( {{\boldsymbol{p}}_{\rm{e}}} \right)}}{{\partial {\theta _{{\rm{e}},{l_2}}}}}} \right)} \right\} \nonumber \\
&= 2{\mathop{\rm Re}\nolimits} \left\{ {{{\bf{C}}_{\rm{{{n^a}}}}}\left( {m,n} \right)} \right\},
\end{align}
where $m = {L}\left( {{l_1} - 1} \right) + {l_1}$, $n = {L}\left( {{l_2} - 1} \right) + {l_2}$, the associated covariance matrix is written as ${{\bf{C}}_{{{{\rm{n}^a}}}}} =\frac{1}{{{\sigma ^2}}}\left( {{\bf{\tilde A}}_1^{\mathsf T} \otimes {{\left( {{\bf{C}} \odot {\bf{B}}} \right)}^{\mathsf T}}} \right)\left( {{\bf{\tilde A}}_1^* \otimes {{\left( {{\bf{C}} \odot {\bf{B}}} \right)}^*}} \right)$, and ${{{\bf{\tilde A}}}_1} = 
\left[ {{{{\bf{\mathord{\buildrel{\lower3pt\hbox{$\scriptscriptstyle\frown$}} 
\over a} }}}_1},{{{\bf{\mathord{\buildrel{\lower3pt\hbox{$\scriptscriptstyle\frown$}} 
\over a} }}}_2}, \ldots, {{{\bf{\mathord{\buildrel{\lower3pt\hbox{$\scriptscriptstyle\frown$}} 
\over a} }}}_L}} \right]$, the $l$-th vector that makes up ${{{\bf{\tilde A}}}_1}$ is expressed as
\begin{align*}
{{{\bf{\mathord{\buildrel{\lower3pt\hbox{$\scriptscriptstyle\frown$}} 
\over a} }}}_l}(i,j) = {\bf{\tilde V}}{{\bf{a}}_{{\rm{IRS,}}l}}(i,j)\left\{ {\frac{{j2\pi \left( {i - 1} \right)cd{\tau_l}\cos ({\theta _{{\rm{e}},l}})\sin ({ \phi _{{\rm{a}},l}})}}{{{\lambda _{\rm{c}}}{u_{i,j}}}}} \right.\notag\\
 \left. {{\kern 1pt}  - \frac{{j2\pi \left( {j - 1} \right)d{u_l}\sin ({\theta _{{\rm{e}},l}})}}{{{\lambda _{\rm{c}}}{u_{i,j}}}}} \right\},
\end{align*}
$1 \le i \le {N_{\rm{y}}}, 1 \le j \le {N_{\rm{z}}}$. The remaining main diagonal submatrices of the FIM, ${{\bf{J}}_{22}}$, ${{\bf{J}}_{33}}$, ${{\bf{J}}_{44}}$, ${{\bf{J}}_{55}}$ are shown in Appendix~\ref{Appendix B}.

Similarly, the $\left ( l_1,l_2 \right )$-th element of the nondiagonal submatrix of the FIM with respect to ${{\bm{\uptheta }}_{\rm{e}}}$ and ${\bm{\upphi} _{\rm{a}}}$ is expressed as
\begin{align}
{{\bf{J}}_{12}}\left( {{l_1},{l_2}} \right) &= \mathbb{E}\left\{ {{{\left( {\frac{{\partial L_{{\rm{o}}}\left( {{\boldsymbol{p}}_{\rm{e}}} \right)}}{{\partial {\theta _{{\rm{e}},{l_1}}}}}} \right)}^*}\left( {\frac{{\partial L_{{\rm{o}}}\left( {{\boldsymbol{p}}_{\rm{e}}} \right)}}{{\partial { \phi _{{\rm{a}},{l_2}}}}}} \right)} \right\} \nonumber \\
&= 2{\mathop{\rm Re}\nolimits} \left\{ {{{\bf{C}}_{\rm{{{n^a},{n^b}}}}}\left( {m,n} \right)} \right\}.
\end{align}
where 
\begin{align*}
{{\bf{C}}_{{\rm{n^a}},{{\rm{n^c}}}}} = \frac{1}{{{\sigma ^4}}}\left( {{\bf{\tilde A}}_1^{\mathsf T} \otimes {{\left( {{\bf{C}} \odot {\bf{B}}} \right)}^{\mathsf T}}} \right){{\bf{C}}_{1,3}}\left( {{{{\bf{\tilde B}}}^*} \otimes {{\left( {{\bf{C}} \odot {\bf{A}}} \right)}^*}} \right),
\end{align*}
${{\bf{C}}_{1,3}} = \mathbb{E}\left\{ {{\rm vec}\left( {{\bf{W}}_{(1)}^{\mathsf H}} \right){\rm vec}\left( {{{\bf{W}}_{(1)}}} \right)} \right\}$, and  ${\bf{\tilde B}} = \left[ {{{{\bf{\bar b}}}_1},{{{\bf{\bar b}}}_2}, \ldots ,{{{\bf{\bar b}}}_L}} \right]$, with ${{\bf{\bar b}}_l}(n) = {{\bf{F}}^{\mathsf H}}{{\bf{a}}_{{\rm{ue}}}}(n)\frac{{ - j2\pi (n - 1)d\cos ({\psi _l})}}{{{\lambda _{\rm{c}}}}}$, $ 1 \le n \le {N_{\rm{t}}}$.
The remaining nondiagonal submatrices of the FIM are provided in Appendix~\ref{Appendix B}. After calculating the components of the FIM, the CRLB can be obtained as
\begin{equation}
{\rm{CRLB}}\left( {{\bf{p}}_{\rm{e}}} \right) = \mathrm{tr}\left [ {{{\bf{J}}^{ - 1}}\left( {{\bf{p}}_{\rm{e}}} \right)} \right ],
\end{equation}
where $\mathrm{tr}\left [ \cdot  \right ]$ denotes the trace operation.

 { In our derivation, the final CRLB, as shown in (31) to (34), fully accounts for the covariance between decoupled parameters (such as distance and angles) after tensor modeling. Since the distance parameters appear in both the IRS array factor matrix and the delay factor matrix, when computing the derivatives of the distance parameter terms in the FIM, we apply the multivariate chain rule to differentiate the distance parameters in both the IRS array factor matrix and the delay factor matrix separately. This enables us to obtain the complete information corresponding to the distance parameters in the FIM matrix blocks. Based on the combination principle of the FIM, we consolidate all channel parameters, including both angle and distance parameters, into a large parameter vector. Using the given log-likelihood function, we compute the FIM corresponding to this parameter vector. This matrix naturally captures the fundamental statistical relationships in the model: the diagonal blocks reflect the correlation of each individual parameter, while the off-diagonal blocks related to the angle and distance parameters explicitly quantify the covariance between the angle and distance parameters.}

{ For the proposed estimator, $\hat{\bf p}$ is the parameter vector to be estimated, while ${\bf p}$ represents the true parameter vector. The biased MSE can be decomposed as ${\text {MSE}}=\mathrm{tr}\left [ {{{\bf{C}}_{ {\rm p}}}} \right ]+\left \| {{\bf b}_{\rm p}} \right \|^2 $, where ${{{\bf{C}}_{ \rm{p}}}}  = \mathbb{E} \left\{[\hat{\bf p} - \mathbb{E}\{\hat{\bf p}\}][\hat {\bf p} - \mathbb{E}\{\hat{\bf p}\}] ^{*}\right\}$ represents the covariance matrix, reflecting the statistical fluctuations of the estimator, and ${{\bf b}_{\rm p}}=\mathbb{E}\{\hat{\bf p}\}]-\bf p$ represents the systematic deviation of the estimator~\cite{CR1}. The randomness of the statistical fluctuation term $\mathrm{tr}\left [ {{{\bf{C}}_{ {\rm p}}}} \right ]$ originates from noise, and its magnitude is therefore governed by the noise variance, typically exhibiting a positive correlation with it~\cite{r28}. In other words, this term mainly determines the slope of the estimation error curve as a function of the signal-to-noise ratio (SNR). In general, the MSE of the proposed estimator is greater than or equal to the CRLB~\cite{CR3}. The fundamental reason for the performance gap between them lies in the inherent systematic bias. From the above MSE decomposition, it follows that reducing the systematic bias term $\left \| {{\bf b}_{\rm p}} \right \|^2$ is an effective way to enable the proposed estimator to approach the CRLB.  Specifically, the proposed estimator adopts a multistage recursive estimation framework based on truncated SVD, where the primary sources of systematic bias include information loss caused by truncated SVD and quantization errors introduced by the angular codebook during angle parameter extraction. To mitigate the impact of systematic bias, a high-resolution codebook can be employed to enrich prior information and minimize quantization errors. Simultaneously, leveraging the low-rank sparsity of the channel helps suppress  weak-path components, thereby reducing information loss during truncated SVD. Fundamentally, both mechanisms act to suppress systematic bias, thereby narrowing the performance gap between the proposed estimator and the CRLB. 
   } 

\begin{table}[htbp]
	\centering
	\footnotesize
	\caption{Computational Complexity Comparison}
	\label{tab:complexity_comparison}
	\renewcommand{\arraystretch}{1.5} %
	\begin{tabularx}{\columnwidth}{|>{\centering\arraybackslash}m{2.5cm}|>{\centering\arraybackslash}X|}
		\hline
		{\textbf{Algorithm}} & {\textbf{Computational Complexity}} \\
		\hline
		{Proposed} & 
		{$\mathcal{O}(PT_{a}^{2}LQ^{2}+QLG_{z}G_{y}+PT_{a}^{2}L^{2}+T_{a}G_{u})$} \\
		\hline
		{PSOMP} & 
		{$\mathcal{O}(QPT_{a}L(G_{1}+G_{2}+G_{3})+L^{4})$} \\
		\hline
		\raisebox{-1.2ex}{{PF-RCE}} &
		{$\mathcal{O}(QPT_{a}L(G_{1}+G_{2}+G_{3})+Q^{2}G_{1}+T_{a}^{2}G_{2}+P^{2}G_{3}+L^{3})$} \\
		\hline
	\end{tabularx}
\end{table}

\subsection{Complexity Analysis}
This section examines the time complexity of the proposed algorithm, with its key components outlined in Algorithm~\ref{alg:alg1}. The complexity of the algorithm is dominated by the SVD, EVD, factor matrix estimation, and channel parameter extraction steps. The corresponding complexities for each step are ${\cal O}\left( {PT_{{\rm{a}}} {Q^2} L} \right)$, ${\cal O}\left( {{L^3}} \right)$, ${\cal O}(PL+PT_{{\rm{a}}}^2 L^2+{PT_{{\rm{a}}}^2 L\left( {Q + L} \right)} )$ and ${\cal O}\left( {QLG_{\rm{z}}G_{\rm{y}}}+{T_{{\rm{a}}}G_{\rm{u}}} \right)$, respectively.
Therefore, the total complexity is $\mathcal{O}\left(PT_\text{a}^2LQ^2+QLG_{\rm{z}}G_{\rm{y}}+PT_\text{a}^2L^2+T_\text{a}G_{\rm{u}}\right)$.

Both polar domain simultaneous orthogonal matching pursuit (PSOMP) algorithm~\cite{r10} and polar-domain frequency-dependent IRS-assisted channel estimation (PF-RCE) algorithm~\cite{rWu} are utilized as benchmarks.
According to \cite{r10}, the total computational complexity of PSOMP algorithm is ${\cal O}\left( {QPT_{{\rm{a}}}L\left( {{G_1} + {G_2} + {G_3}} \right) + {L^4}} \right)$. According to \cite{rWu}, the PF-RCE algorithm yields the complexity of ${\cal O}\left( {QPT_{{\rm{a}}}L\left( {{G_1} + {G_2} + {G_3}} \right) + {Q^2}{G_1} + {T_{{\rm{a}}}^2 }{G_2} + {P^2}{G_3} + {L^3}} \right)$, where ${{G_1}}$, ${{G_2}}$ and ${{G_3}}$ represent the grid dimensions utilized in the estimation algorithms.

\begin{figure}[htbp]
\centering
\includegraphics[width=2.9in]{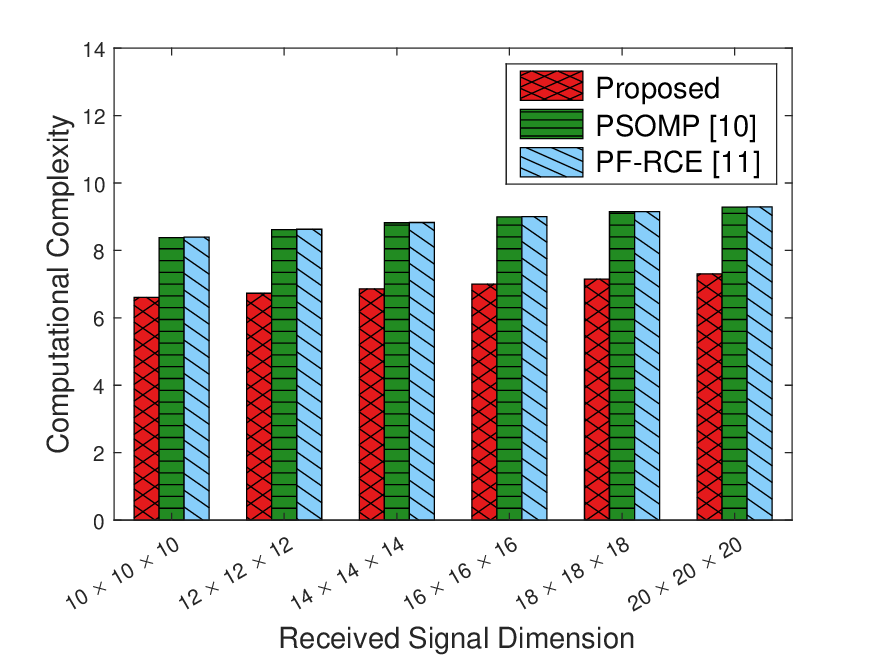}
\caption{Comparison of the computational complexity (logarithmic with base 10) performance for three different methods as  the received signal dimension $Q\times P\times T_{\rm{a}}$ varies.}
\label{fig.4}
\end{figure}

In Fig. \ref{fig.4}, we compare the computational complexity of the proposed algorithm with benchmarks. 
We set ${{G_1=58746}}$, ${{G_2=500}}$, ${{G_3=1000}}$, $G_{\rm{z}}=G_{\rm{y}}=300$, $G_{\rm{u}}=2000$ and $Q \times P \times T_{{\rm{a}}}$ to values ranging from 10 to 20. It is observed that the proposed algorithm demonstrates significantly lower computational complexity compared to PSOMP and PF-RCE.

\section{Simulation Results}\label{V}
This section conducts several simulations to evaluate the proposed algorithm, where the metric of the normalized mean square error (NMSE) with the signal-to-noise ratio (SNR)~\cite{r33} is adopted, i.e., 
\begin{equation*}
\begin{aligned}
    \text{NMSE}\left( p \right) = \frac{{\left\| {p - \hat p} \right\|_2^2}}{{\left\| p \right\|_2^2}} \quad  \text{and} \quad 
    \text{SNR}= \frac{{\left\| {{\cal Y} - {\cal N}} \right\|_F^2}}{{\left\| {\cal N} \right\|_F^2}},
\end{aligned}
\end{equation*}
where $p=\left\{ {{\theta _{{\rm{e}},l}},{ \phi _{{\rm{a}},l}},{\tau_l},{{ \psi _l}},{ \gamma _l}} \right\}_{l=1}^{L}$ denotes the true channel parameters and $\hat{p} =\left\{ {{\hat\theta _{{\rm{e}},l}},{\hat \phi _{{\rm{a}},l}},{\hat \tau_l},{{\hat \psi _l}},{\hat \gamma _l}} \right\}_{l = 1}^{{L}}$ denotes the estimated channel parameters.

The simulation setup in the considered scenario is as follows: ${N_{\rm{r}}}=512$, ${N_{\rm{b}}}=64$, ${N_{\rm{t}}}=64$, ${L} = 4$. Moreover, the carrier frequency, bandwidth, and number of subcarriers are set to ${f_{\rm{c}}} = 100 \;\mathrm{GHz}$, ${f_{\mathrm{s}}} = 320\;\mathrm{MHz}$, and $P = 16$, respectively. The uplink pilot structure adopts $T_{{\rm{a}}} = 16$ time frames, each with $Q = 16$ time slots. The path gain follows a circular Gaussian distribution. All path directions are randomly generated from $(0,2\pi)$. According to~\cite{r12}, the Rayleigh distance in the considered system is given by $6.24\;m$. {To obtain stable experimental results, we conducted 500 Monte Carlo experiments}.

\begin{figure}[t]
    \centering
	\subfloat[Azimuth and Elevation]{
    \includegraphics[width=0.24\textwidth]{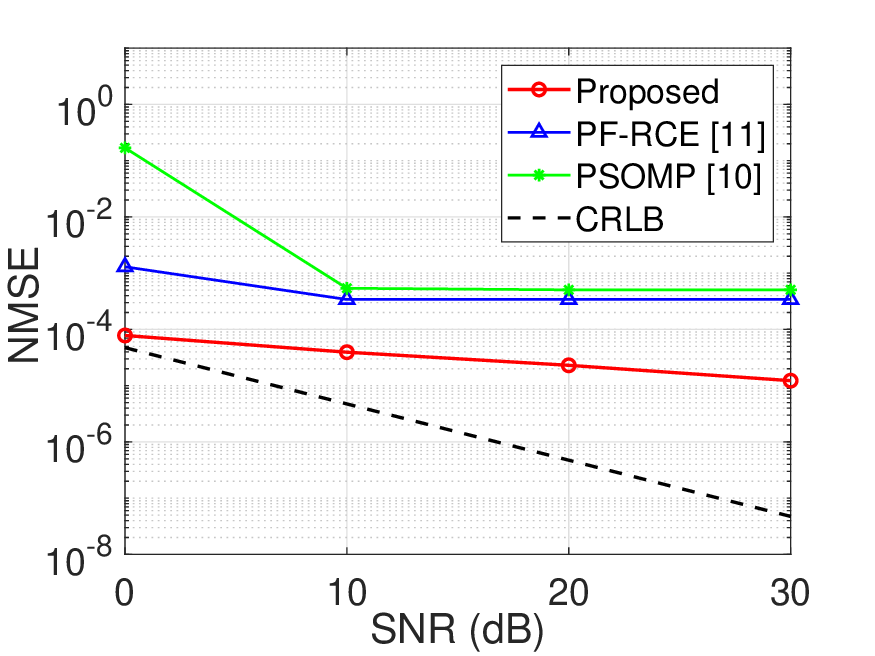}}
    \subfloat[AoD]{
    \includegraphics[width=0.24\textwidth]{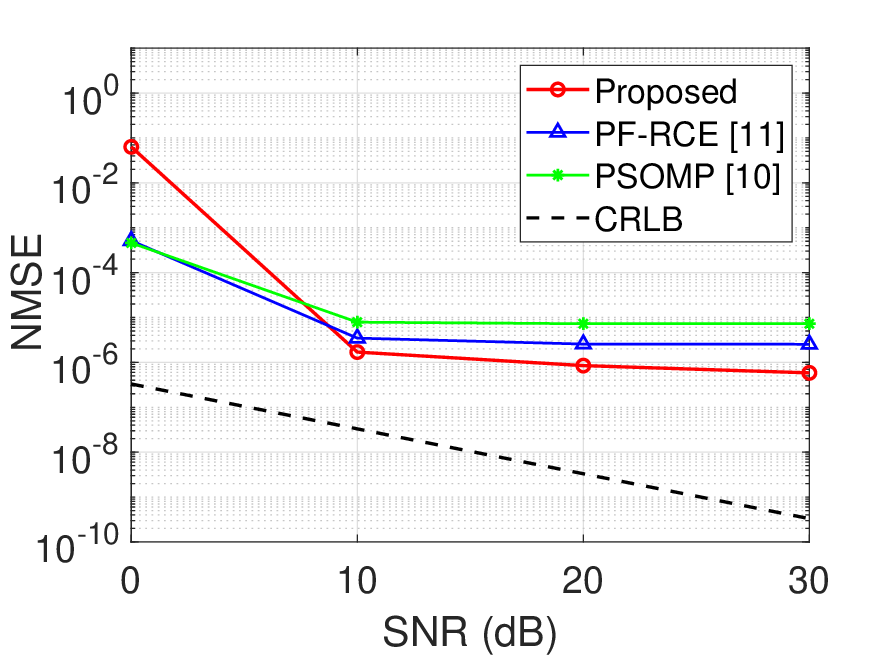}}\\
    \subfloat[Delay]{
    \includegraphics[width=0.24\textwidth]{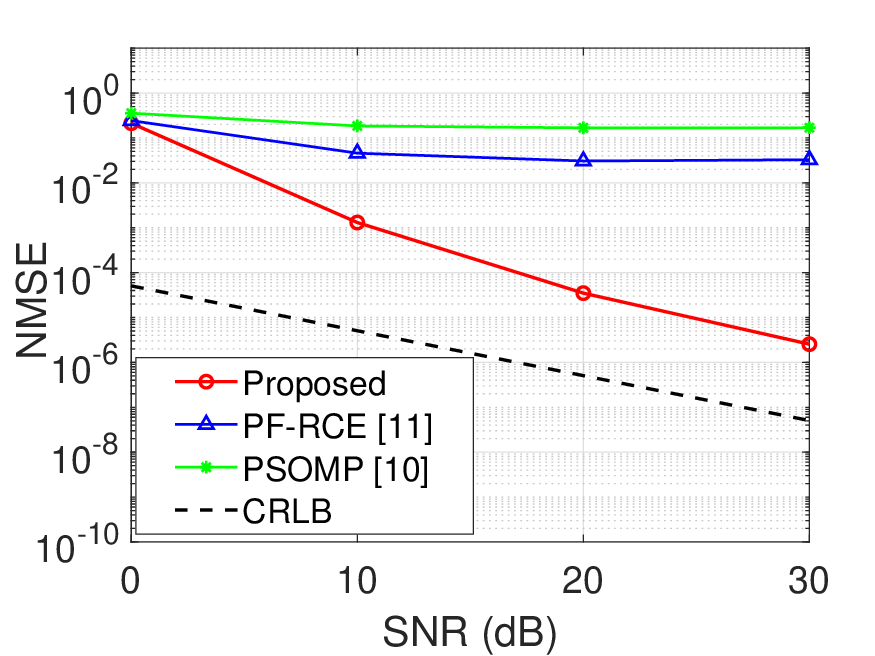}}  
    \subfloat[Gain]{
    \includegraphics[width=0.24\textwidth]{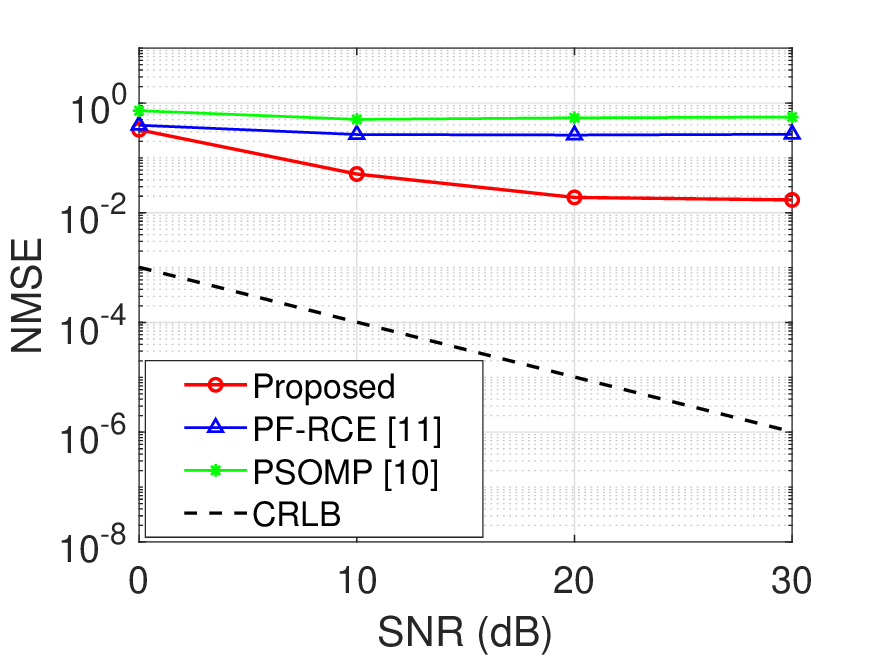}}
\caption{NMSE performance of each algorithm and the CRLB as the SNR varies, where the distance between the user and the IRS is chosen over a range of (1m, 6m).}
\label{fig.5}
\end{figure}

\begin{figure}[t]
    \centering
	\subfloat[Azimuth and Elevation]{
    \includegraphics[width=0.24\textwidth]{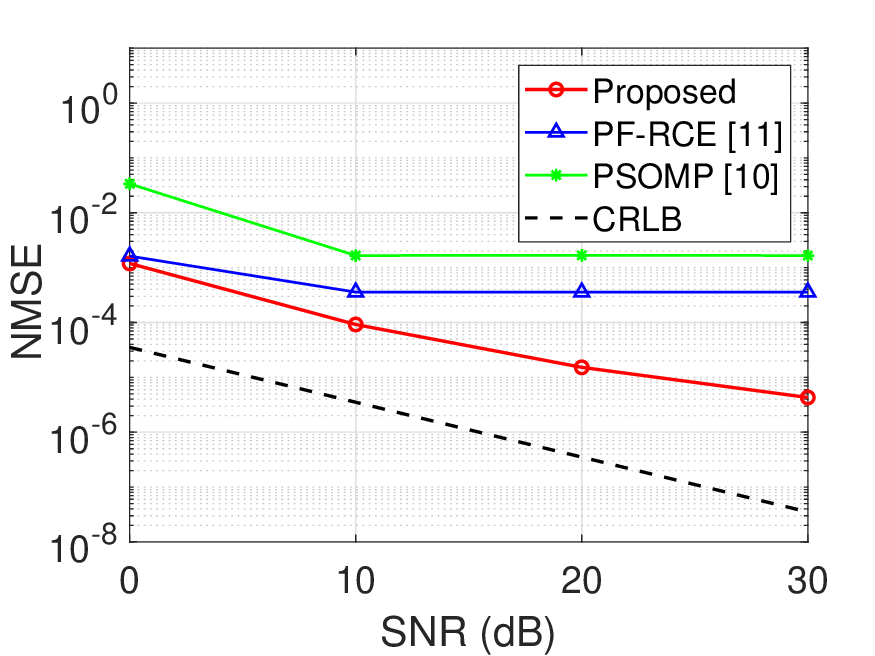}}
    \subfloat[AoD]{
    \includegraphics[width=0.24\textwidth]{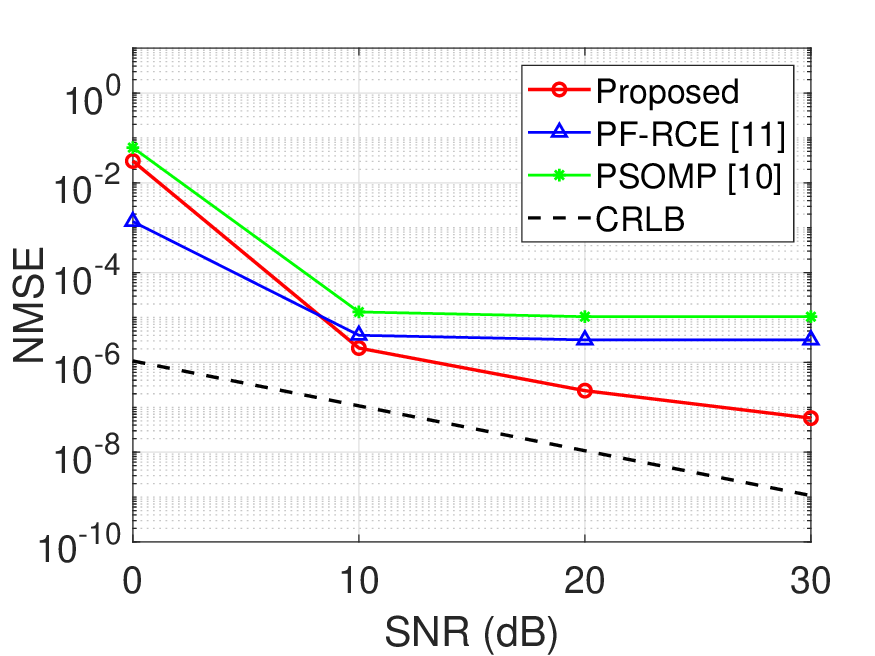}}\\
    \subfloat[Delay]{
    \includegraphics[width=0.24\textwidth]{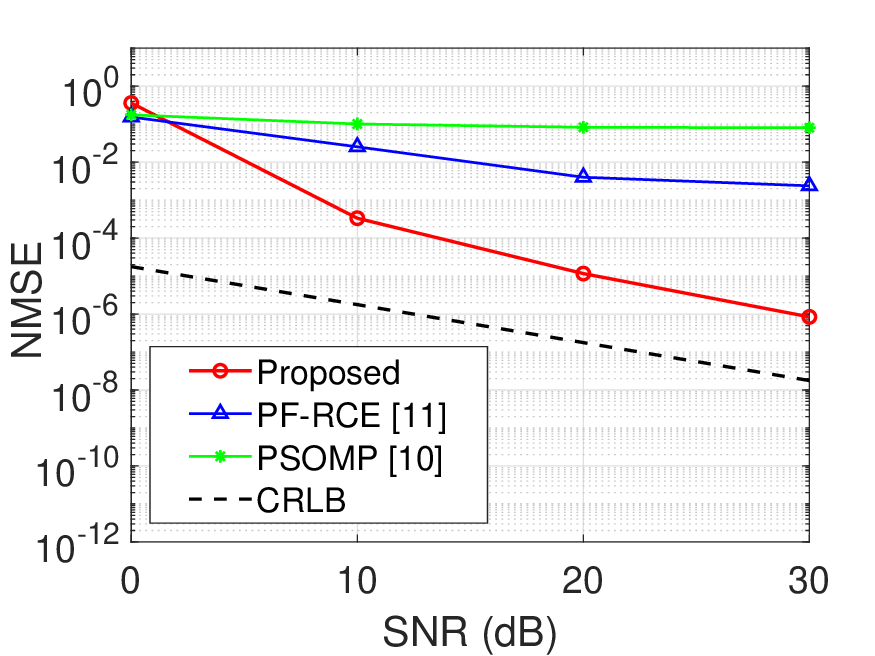}}
    \subfloat[Gain]{
    \includegraphics[width=0.24\textwidth]{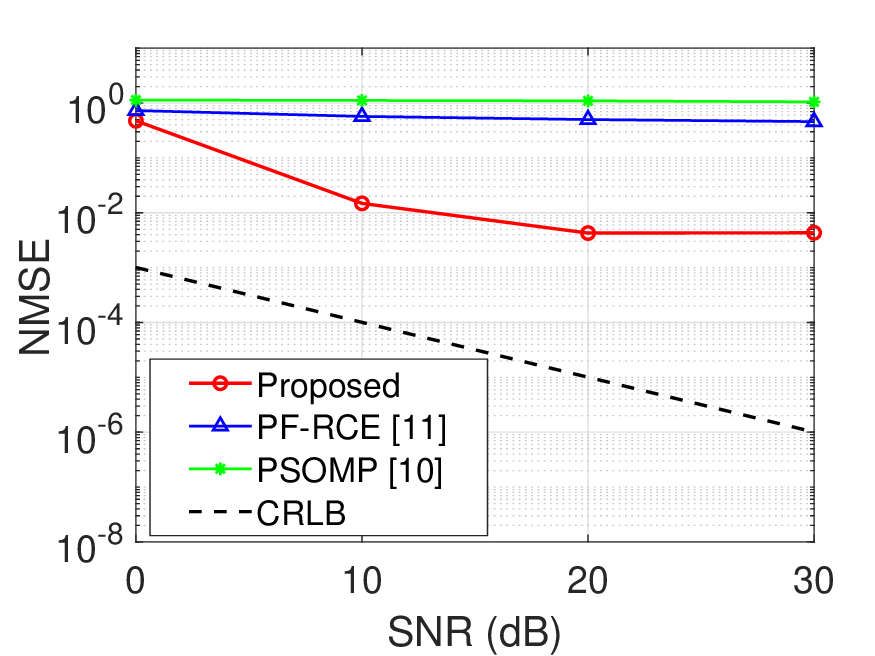}}
\caption{NMSE performance of each algorithm and the CRLB as SNR varies, where the distance between the user and the IRS is chosen over a range of (15m, 20m).}
\label{fig.6}
\end{figure}

To assess the impact of the distance on the estimation performance, We verify the channel estimation methods on two selected ranges (1m, 6m) in Fig.~\ref{fig.5} and (15m, 20m) in Fig.~\ref{fig.6}. 
The estimation performance of azimuth/elevation at the IRS is shown in Fig.~\ref{fig.5}(a). It is seen that NMSEs of all algorithms exhibit a declining trend as SNR increases. The proposed algorithm is closer to the CRLB, which confirms its effectiveness. Fig.~\ref{fig.5}(b) shows the AoD estimation of the
UE, where it is observed that at low SNR, our method performs worse than the CS method. This is because the single parameter detection codebook in the correlation-based method is more sensitive to noise. However, as the SNR increases, the CS method continues to accumulate errors during the iterative process, which reduces the estimation performance. In contrast, our method can efficiently and accurately estimate the AoD.

The estimation results of the delay and complex path gain
are illustrated in Figs.~\ref{fig.5}(c) and (d), respectively. As SNR increases, our proposed method demonstrates excellent performance. In Fig.~\ref{fig.5}(d), the NMSE performance of the path gain deviates slightly from the CRLB, due to error accumulation
from the sequential estimation of angle and delay parameters.

Comparing the channel estimation performance between Figs.~\ref{fig.5} and \ref{fig.6}, our algorithm consistently outperforms benchmark methods across varying distances, demonstrating robust and reliable estimation accuracy.

\begin{figure}[t]
   \centering
	\subfloat[]{
    \includegraphics[width=0.24\textwidth]{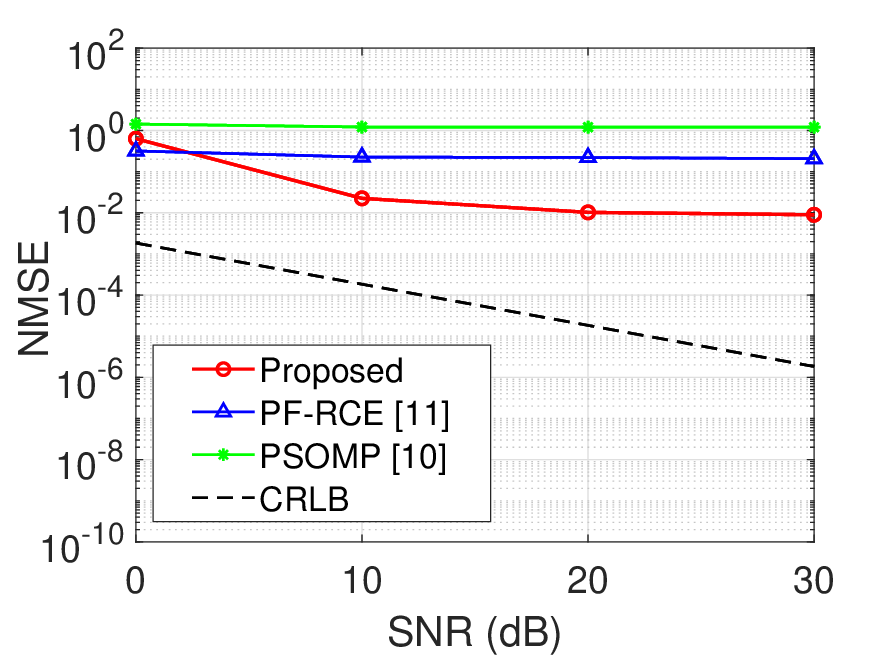}}
    \subfloat[]{
    \includegraphics[width=0.24\textwidth]{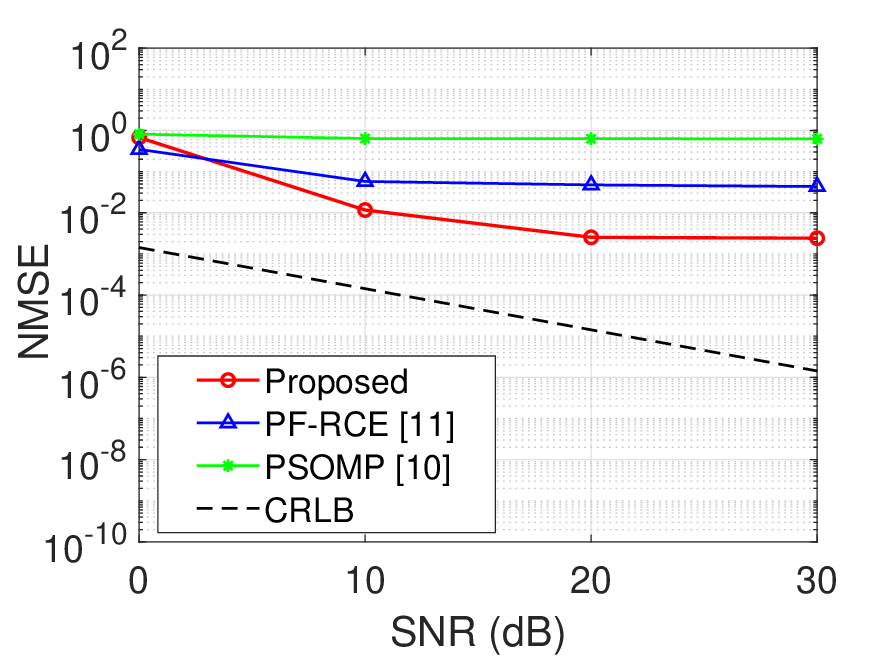}}
\caption{NMSE performance of the proposed algorithm and the comparison algorithms for the overall channel variation with the SNR, where (a) denotes the  distance selected from (1m, 6m) and (b) is the distance selected from (15m, 20m).}
\label{fig.7}
\end{figure}

Fig.~\ref{fig.7} presents the NMSE performance of the overall channel, integrating all estimated channel parameters as described in \eqref{eq_80}. {The expression for the overall channel is derived in Appendix~B}. {The two subplots in Fig.~\ref{fig.7} show the NMSE performance at two different distances, representing the far-field and NF conditions}. The proposed algorithm exhibits good performance even under different distances. Fig.~\ref{fig.7}(a) compares the NMSE performance of various algorithms in a NF scenario. By calculating the average NMSE difference at each SNR between our method and the PF-RCE algorithm (the leading CS-based approach), we observe that our method delivers an 8.5 dB improvement, underscoring its enhanced accuracy. { In the low-SNR regime of 0 dB to 10 dB, the NMSE of the proposed estimator decreases with a slope nearly identical to the CRLB, despite a persistent performance gap. In this region, the estimation error is predominantly governed by noise; thus, the noise-induced statistical fluctuations diminish significantly as the SNR increases, while the systematic bias remains relatively constant. Since the CRLB is determined solely by noise statistics, both curves exhibit similar descending trends. The gap between the estimator and the CRLB lies in the presence of inherent system bias, such as quantization errors introduced by the angular codebook and information loss caused by truncated SVD. When the SNR enters the high-SNR regime (above 10 dB), the influence of noise on the SNR-dependent variation of the NMSE gradually diminishes, and the system bias becomes the dominant error source. Consequently, the NMSE curve flattens as the estimator hits an error floor, whereas the CRLB continues to descend with noise reduction. This divergence leads to a widening performance gap in the high-SNR region.}

\begin{figure}[t]
	\centering
   \includegraphics[width=0.5\textwidth]{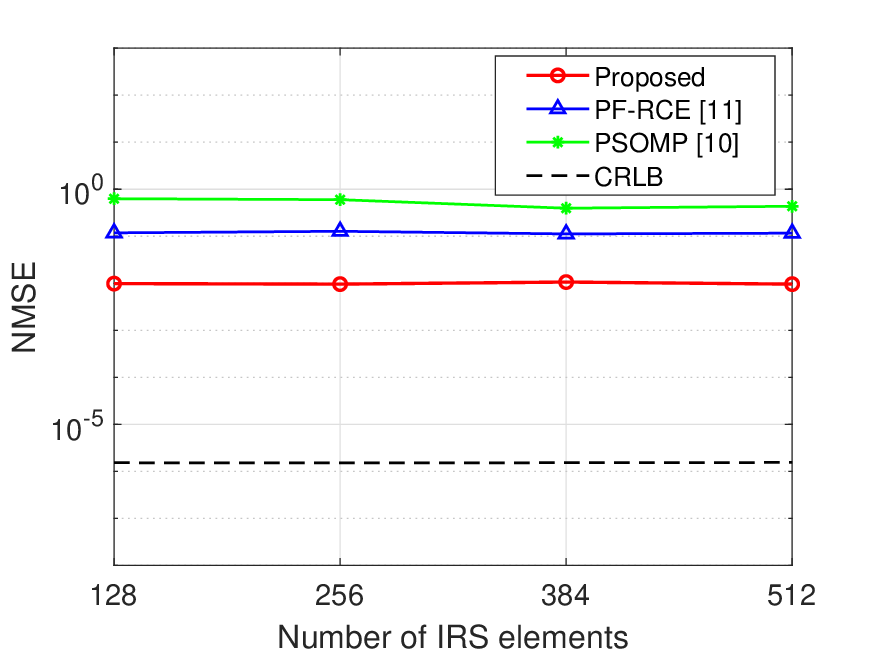}
	\caption{{The overall channel NMSE performance of the proposed algorithm and comparison algorithms under varying IRS elements, with SNR =30 dB.}}
	\label{fig.8}
\end{figure}

{In Fig.~\ref {fig.8}, we present the channel estimation performance (NMSE) of the proposed algorithm and the comparison algorithms as a function of the number of IRS elements. The results show that the NMSE does not vary significantly with changes in the number of IRS elements. This is primarily because, in our method, the information from all IRS elements is compressed and mapped into a low-dimensional subspace when constructing the IRS array factor matrix (as shown in the structure of the factor matrix $\bf A$ in Eq. (12)). Similarly, in the derivation of the CRLB, the information related to the IRS elements in the FIM submatrix (as in Eq. (32)) is also mapped to a low-dimensional subspace. As a result, the NMSE performance under the CRLB does not show significant variation with changes in the number of IRS elements.}

\begin{figure}[t]
	\centering
   \includegraphics[width=0.5\textwidth]{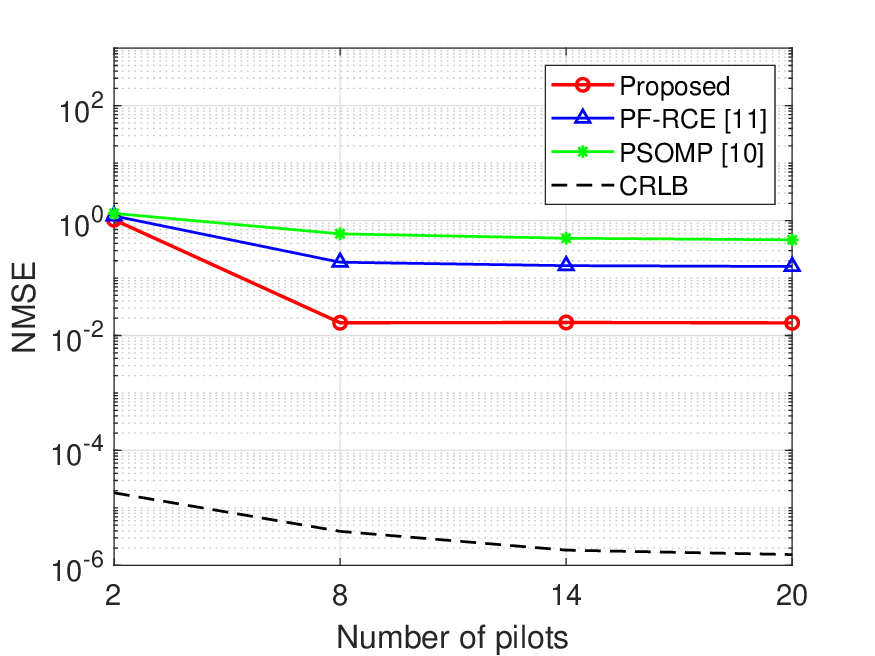}
	\caption{{{The overall channel NMSE performance of the proposed algorithm and comparison algorithms under varying pilot numbers, with SNR =30 dB.}}}
	\label{fig.9}
\end{figure}

{In Fig. ~\ref {fig.9}, we present the simulation results of the proposed algorithm and the comparison algorithms with varying pilot numbers. According to the channel estimation training design discussed earlier, the number of pilots is denoted as $T_{\rm{a}}$. The experimental results clearly show that as the number of pilots increases, the sample size of the observed data expands, which leads to a continuous improvement in the NMSE performance of channel estimation. After the pilot number reaches a certain scale, the NMSE curve of the proposed algorithm gradually stabilizes. Despite the saturation phenomenon, the proposed algorithm outperforms other comparison schemes across the entire pilot range, demonstrating the effectiveness and robustness of its structure.} 

\begin{figure}[t]
	\centering
   \includegraphics[width=0.5\textwidth]{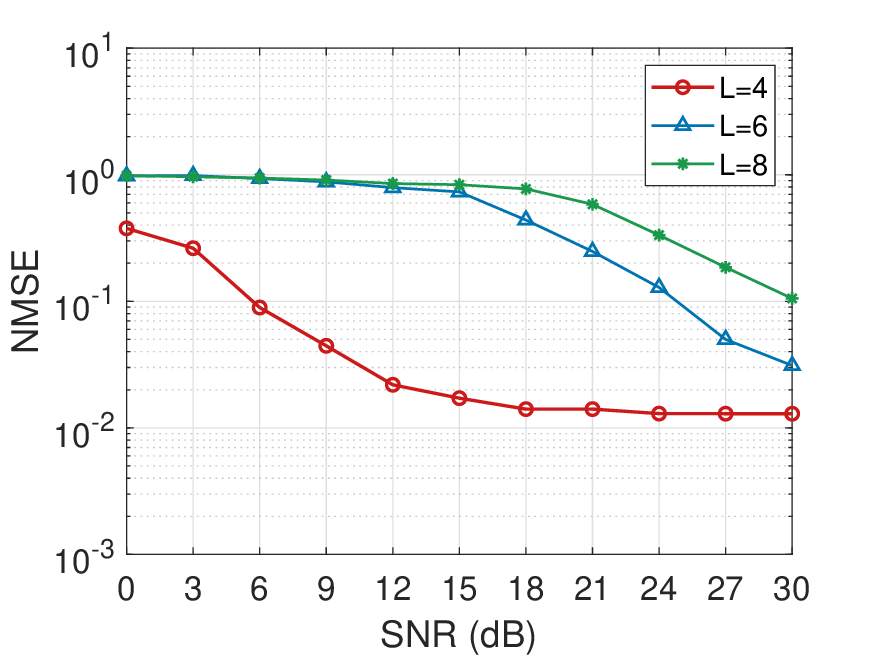}
	\caption{{The NMSE of the proposed algorithm estimates with respect to SNR, with three different curves corresponding to different path values.}}
	\label{fig.10}
\end{figure}

{Fig.~\ref {fig.10} illustrates the NMSE performance of the proposed algorithm under different numbers of propagation paths. When the number of paths is small, the boundary between the signal and noise subspaces is clearly distinguishable. In this case, the first step of the proposed method, SVD, can effectively separate the signal and noise subspaces, thereby improving output quality and enhancing estimation accuracy. As shown in Fig.~\ref {fig.10}, the NMSE performance for $L=4$ is significantly better than that for 
$L=6$ and $L=8$. On the other hand, as the number of paths increases, the decay rate of the singular values associated with the signal subspace becomes slower, leading to stronger overlap between the signal and noise components during the SVD process. This overlap introduces additional estimation errors and limits further performance improvement. As illustrated in Fig.~\ref {fig.10}, the NMSE curves for $L = 6$ and $L = 8$ exhibit degraded performance in the low-SNR regime. Moreover, since the difference in the degree of mixing between the signal and noise subspaces is small under this condition, the strong noise masks these subtle distinctions, resulting in similar NMSE performance. As the SNR increases, the influence of noise diminishes, and the separation between the signal and noise subspaces becomes more distinct with fewer paths, yielding higher estimation accuracy. Consequently, the NMSE differences among different path numbers become more pronounced at high SNRs.} 

\section{Conclusion and Future Directions}\label{VI}
In this paper, we addressed the channel estimation challenge in NF IRS communication systems by developing a harmonic processing-inspired tensor modalization framework. Our approach modeled the high-dimensional NF channel as a tensor, decomposing it into independent factor matrices that decouple distance and angle parameters, akin to chords in musical harmonic analysis. By exploiting the Vandermonde structure and low-rank properties of these matrices, we first estimated distance parameters with high precision. Subsequently, a compact, distance-dependent codebook was designed to achieve efficient angle parameter extraction via correlation detection. This codebook design significantly reduced search complexity compared to traditional polar-domain methods. Theoretical analysis, including the CRLB derivation, validated the robustness of our estimators. Simulation results confirmed the superior accuracy and computational efficiency of the proposed method.     

{Future research can consider channel estimation in NF wideband scenarios, where both NF effects and beam squint phenomena should be taken into account. Moreover, subsequent studies could further investigate the structural characteristics of NF IRS arrays, for instance, exploiting the symmetry of the NF array response with respect to the element index can improve estimation efficiency.
}

\appendix

\subsection{Parametric Derivation}\label{Appendix A}
We first derive the result of \eqref{eq_31}. According to the matrix derivation, we have
\begin{equation}
\frac{{\partial L_{{\rm{o}}}\left( {{\bf{p}}_{\rm{e}}} \right)}}{{\partial {\theta _{{\rm{e}},l}}}} = \mathrm{tr}\left\{ {{{\left( {\frac{{\partial L_{{\rm{o}}}\left( {{\bf{p}}_{\rm{e}}} \right)}}{{\partial {\bf{A}}}}} \right)}^{\mathsf T}}\frac{{\partial {\bf{A}}}}{{\partial {\theta _{{\rm{e}},l}}}} + {{\left( {\frac{{\partial L_{{\rm{o}}}\left( {{\bf{p}}_{\rm{e}}} \right)}}{{\partial {{\bf{A}}^*}}}} \right)}^{\mathsf T}}\frac{{\partial {{\bf{A}}^*}}}{{\partial {\theta_{{\rm{e}},l}}}}} \right\}.
\label{eq_37}
\end{equation}
In Eqn.~\eqref{eq_37}, 
\begin{align}
\frac{{\partial L_{{\rm{o}}}\left( {{\bf{p}}_{\rm{e}}} \right)}}{{\partial {\bf{A}}}} &= \frac{1}{{{\sigma ^2}}}{\left( {{\bf{Y}}_{(1)}^{\mathsf T} - \left( {{\bf{C}} \odot {\bf{B}}} \right){{\bf{A}}^{\mathsf T}}} \right)^{\mathsf H}}\left( {{\bf{C}} \odot {\bf{B}}} \right), \label{eq_38}\\
\frac{{\partial L_{{\rm{o}}}\left( {{\bf{p}}_{\rm{e}}} \right)}}{{\partial {{\bf{A}}^*}}} &= {\left( {\frac{{\partial L_{{\rm{o}}}\left( {{\bf{p}}_{\rm{e}}} \right)}}{{\partial {\bf{A}}}}} \right)^*}, \label{eq_39} \\
\frac{{\partial {\bf{A}}}}{{\partial {\theta _{{\rm{e}},l}}}} &= \left[ {0, \ldots, {{\mathord{\buildrel{\lower3pt\hbox{$\scriptscriptstyle\frown$}} 
\over {\boldsymbol{a}}} }_l}, \ldots , 0} \right],
\label{eq_40}
\end{align}
where
\begin{align*}
{{{\bf{\mathord{\buildrel{\lower3pt\hbox{$\scriptscriptstyle\frown$}} 
\over a} }}}_l}(i,j) = {\bf{\tilde V}}{{\bf{a}}_{{\rm{IRS,}}l}}(i,j)\left\{ {\frac{{j2\pi \left( {i - 1} \right)cd{\tau_l}\cos ({\theta _{{\rm{e}},l}})\sin ({ \phi _{{\rm{a}},l}})}}{{{\lambda _{\rm{c}}}{u_{i,j}}}}} \right.\notag\\
 \left. {{\kern 1pt}  - \frac{{j2\pi \left( {j - 1} \right)d{u_l}\sin ({\theta _{{\rm{e}},l}})}}{{{\lambda _{\rm{c}}}{u_{i,j}}}}} \right\}, \quad 1 \le i \le {N_{\rm{y}}}, 1 \le j \le {N_{\rm{z}}}.
\end{align*}

Substituing Eqs.~\eqref{eq_38}-\eqref{eq_40} into \eqref{eq_37} yields
\begin{align}
&\frac{{\partial L_{{\rm{o}}}\left( {{\boldsymbol{p}}_{\rm{e}}} \right)}}{{\partial {\theta _{{\rm{e}},l}}}} = {\bf{e}}_l^{\mathsf T}\frac{1}{{{\sigma ^2}}}{\left( {{\bf{C}} \odot {\bf{B}}} \right)^{\mathsf T}}{\left( {{\bf{Y}}_{(1)}^{\mathsf T} - \left( {{\bf{C}} \odot {\bf{B}}} \right){{\bf{A}}^{\mathsf T}}} \right)^*}{{{\bf{\mathord{\buildrel{\lower3pt\hbox{$\scriptscriptstyle\frown$}} 
\over a} }}}_l} \nonumber \\
&+ {\bf{e}}_l^{\mathsf T}\frac{1}{{{\sigma ^2}}}{\left( {{\bf{C}} \odot {\bf{B}}} \right)^{\mathsf H}}{\left( {{\bf{Y}}_{(1)}^{\mathsf T} - \left( {{\bf{C}} \odot {\bf{B}}} \right){{\bf{A}}^{\mathsf T}}} \right)^*}{\bf{\mathord{\buildrel{\lower3pt\hbox{$\scriptscriptstyle\frown$}} 
\over a} }}_l^* \nonumber\\
&= 2{\mathop{\rm Re}\nolimits} \left\{ {{\bf{e}}_l^{\mathsf T}\frac{1}{{{\sigma ^2}}}{{\left( {{\bf{C}} \odot {\bf{B}}} \right)}^{\mathsf T}}{{\left( {{\bf{Y}}_{(1)}^{\mathsf T} - \left( {{\bf{C}} \odot {\bf{B}}} \right){{\bf{A}}^{\mathsf T}}} \right)}^*}{{{\bf{\tilde A}}}_1}{{\bf{e}}_l}} \right\},
\end{align}
where ${{{\bf{\tilde A}}}_1} = 
\left[ {{{{\bf{\mathord{\buildrel{\lower3pt\hbox{$\scriptscriptstyle\frown$}} 
\over a} }}}_1},{{{\bf{\mathord{\buildrel{\lower3pt\hbox{$\scriptscriptstyle\frown$}} 
\over a} }}}_2}, \ldots, {{{\bf{\mathord{\buildrel{\lower3pt\hbox{$\scriptscriptstyle\frown$}} 
\over a} }}}_L}} \right]$, ${\mathop{\rm Re}\nolimits} \left\{ { \cdot} \right\}$ denotes the real part of a complex number, and ${{{\boldsymbol{e}}_l}}$ denotes the standard coordinate base with nonzero elements of $l$.
Likewise, we can obtain the partial derivatives of parameters ${\phi_{{\rm{a}},l}}$, ${{\psi _l}}$ and ${\gamma _l}$, as given by 
\begin{align}
\frac{{\partial L_{{\rm{o}}}\left( {{\bf{p}}_{\rm{e}}} \right)}}{{\partial { \phi _{{\rm{a}},l}}}} 
= & \, 2{\mathop{\rm Re}\nolimits} \left\{ {\bf{e}}_l^{\mathsf T}\frac{1}{{{\sigma ^2}}}{{\left( {{\bf{C}} \odot {\bf{B}}} \right)}^{\mathsf T}} \right. \notag\\
& \left. \quad \quad \quad  {{\left( {{\bf{Y}}_{(1)}^{\mathsf T} - \left( {{\bf{C}} \odot {\bf{B}}} \right){{\bf{A}}^{\mathsf T}}} \right)}^*}{{{\bf{\tilde A}}}_2}{{\bf{e}}_l} \right\},
\end{align}

\begin{align}
\frac{\partial L_{{\rm{o}}}\left( {{\bf{p}}_{\rm{e}}} \right)}{\partial \psi_l} 
= & \, 2{\mathop{\rm Re}\nolimits} \left\{ {\bf{e}}_l^{\mathsf T}\frac{1}{\sigma^2}{{\left( {\bf{C}} \odot {\bf{A}} \right)}^{\mathsf T}} \right. \notag\\
& \left.\quad \quad \quad  {{\left( {\bf{Y}}_{(2)}^{\mathsf T} - \left( {\bf{C}} \odot {\bf{A}} \right){\bf{B}}^{\mathsf T} \right)}^*}{\bf{\tilde B}}{\bf{e}}_l \right\},
\end{align}
\begin{align}
\frac{\partial L_{{\rm{o}}}\left( \mathbf{p}_{\rm{e}} \right)}{\partial \gamma_l} 
&= \mathbf{e}_l^{\mathsf T}\frac{1}{\sigma^2}{\left( \mathbf{B} \odot \mathbf{A} \right)^{\mathsf T}}\notag \\
&\quad \quad {\left( \mathbf{Y}_{(3)}^{\mathsf T} - \left( \mathbf{B} \odot \mathbf{A} \right)\mathbf{C}^{\mathsf T} \right)}^*\mathbf{\tilde G}\mathbf{e}_l,
\end{align}
where ${{{\bf{\tilde A}}}_2} = \left[ {{{{\bf{\bar a}}}_1},{{{\bf{\bar a}}}_2}, \ldots ,{{{\bf{\bar a}}}_L}} \right]$, ${{{\bf{\bar a}}}_l}(i,j) = {\bf{\tilde V}}{{\bf{a}}_{{\rm{IRS,}}l}}(i,j)\left\{ {\frac{{j2\pi \left( {j - 1} \right)d{u_l}sin({\theta _{{\rm{e}},l}})\cos ({ \phi _{{\rm{a}},l}})}}{{{\lambda _{\rm{c}}}{u_{i,j}}}}} \right\};$ ${\bf{\tilde B}} = \left[ {{{{\bf{\bar b}}}_1},{{{\bf{\bar b}}}_2}, \ldots ,{{{\bf{\bar b}}}_L}} \right]$, ${{\bf{\bar b}}_l}(n) = {{\bf{F}}^{\mathsf H}}{{\bf{a}}_{{\rm{ue}}}}(n)\frac{{ - j2\pi (n - 1)d\cos ({\psi _l})}}{{{\lambda _{\rm{c}}}}}$, $ 1 \le n \le {N_{\rm{t}}}$; ${\bf{\tilde G}} = \left[ {{{{\bf{\bar g}}}_1},{{{\bf{\bar g}}}_2}, \ldots ,{{{\bf{\bar g}}}_L}} \right]$, ${{\bf{\bar g}}_l} = {\left[ {{e^{ - j2\pi \frac{{{f_s}}}{{{P_0}}}{\tau _l}}}, \ldots ,{e^{ - j2\pi \frac{{{f_s}}}{{{P_0}}}P{\tau _l}}}} \right]^{\mathsf T}}$.

The derivation for ${\tau_l}$ is more complicated because both matrices $\bf{A}$ and $\bf{C}$ contain ${\tau_l}$. Defining ${\bf{R}} = \left( {{\bf{B}} \odot {\bf{A}}} \right){{\bf{C}}^{\mathsf T}}$, we have
\begin{align}
\frac{{\partial L_{{\rm{o}}}\left( {{\bf{p}}_{\rm{e}}} \right)}}{{\partial {\bf{R}}}} &= \frac{1}{{{\sigma ^2}}}{\left( {{\bf{Y}}_{(3)}^{\mathsf T} - {\bf{R}}} \right)^*},\\
\frac{{\partial {\bf{R}}}}{{\partial {\tau_l}}} &= \frac{{\partial {\bf{B}} \odot {\bf{A}}}}{{\partial {\tau_l}}}{{\bf{C}}^{\mathsf T}} + {\bf{B}} \odot {\bf{A}}\frac{{\partial {{\bf{C}}^{\mathsf T}}}}{{\partial {\tau_l}}}. \label{eq_44}
\end{align}
In Eqn.~\eqref{eq_44}, $\frac{{\partial {\bf{B}} \odot {\bf{A}}}}{{\partial {\tau_l}}}$ is given by
\begin{equation}
\frac{{\partial {\bf{B}} \odot {\bf{A}}}}{{\partial {\tau_l}}} = \left[ {0, \ldots, {\bf{c}}_l^{\rm {a}} ,\ldots, 0} \right],\label{eq_45}
\end{equation}
where
\begin{align*}
{\bf{c}}_l^{\rm {a}}(i,j) =& {{{\bf{\tilde a}}}_{{\rm{ue}},l}} \otimes {\bf{\tilde V}}{{\bf{a}}_{{\rm{IRS,}}l}}(i,j)\\
\times & \frac{{ - j2\pi }}{{{\lambda _{\rm{c}}}}}\left( {\left( {\frac{{\left( c^2{{\tau_l} - (i - 1)cdsin({\theta _{{\rm{e}},l}})\sin ({ \phi _{{\rm{a}},l}})} \right)}}{{{u_{i,j}}}}} \right.} \right.\\
- & \left. {\left. {\frac{{(j - 1)d\cos({\theta _{{\rm{e}},l}})}}{{{u_{i,j}}}}} \right) - 1} \right).
\end{align*}
The derivative of ${{\bf{C}}^{\mathsf T}}$ with respect to  ${\tau_l}$ is 
\begin{equation}
\frac{{\partial {{\bf{C}}^{\mathsf T}}}}{{\partial {\tau_l}}} = {\left[ {0, \ldots ,{\bf{c}}_l^{\rm {b}} ,\ldots, 0} \right]^{\mathsf T}}, \label{eq_46}
\end{equation}
where ${\bf{c}}_l^{\rm {b}} = \left[ { - j2\pi {\gamma _l}\frac{{{f_{\rm{s}}}}}{{{P_0}}}{e^{ - j2\pi \frac{{{f_{\rm{s}}}}}{{{P_0}}}{\tau _l}}}, \ldots , - j2\pi {\gamma _l}{e^{ - j2\pi \frac{{{f_{\rm{s}}}}}{{{P_0}}}P{\tau _l}}}} \right]$.
Combining \eqref{eq_45} and \eqref{eq_46} yields
\begin{align}
&\frac{{\partial L_{{\rm{o}}}\left( {{\bf{p}}_{\rm{e}}}\right)}}{{\partial {\tau_l}}} = 2{\mathop{\rm Re}\nolimits} \left\{ {{\bf{e}}_l^{\mathsf T}{{\bf{C}}^{\mathsf T}}\frac{1}{{{\sigma ^2}}}{{\left( {{\bf{Y}}_{(3)}^{\mathsf T} - \left( {{\bf{B}} \odot {\bf{A}}} \right){{\bf{C}}^{\mathsf T}}} \right)}^{\mathsf H}}{\bf{C}}^{\rm {a}}}{\bf{e}}_l \right. \nonumber\\
&\left. { + {\bf{e}}_l^{\mathsf T}{\bf{C}}^{\rm {b}}{{\left( {{\bf{Y}}_{(3)}^{\mathsf T} - \left( {{\bf{B}} \odot {\bf{A}}} \right){{\bf{C}}^{\mathsf T}}} \right)}^{\mathsf H}}\left( {{\bf{B}} \odot {\bf{A}}} \right){{\bf{e}}_l}} \right\},
\label{eq_49}
\end{align}
where ${\bf{C}}^{\rm {a}} = \left[ {{\bf{c}}_1^{\rm {a}} ,\ldots, {\boldsymbol{c}}_{{L}}^{\rm {a}}} \right]$ and ${\bf{C}}^{\rm {b}} = {\left[ {{\bf{c}}_1^{\rm {b}} ,\ldots, {\bf{c}}_{{L}}^{\rm {b}}} \right]^{\mathsf T}}$.

\subsection{FIM Elements} \label{Appendix B}
To determine different submatrices of the FIM matrix, we first derive the diagonal submatrix of the FIM in relation to the variable ${\tau}$ as 
\begin{align}
&{{\bf{J}}_{44}}\left( {{l_1},{l_2}} \right)=\mathbb{E}\left\{ {{{\left( {\frac{{\partial L_{{\rm{o}}}\left( {{\bf{p}}_{\rm{e}}} \right)}}{{\partial {\tau_{{l_1}}}}}} \right)}^*}\left( {\frac{{\partial L_{{\rm{o}}}\left( {{\bf{p}}_{\rm{e}}} \right)}}{{\partial {\tau_{{l_2}}}}}} \right)} \right\}\notag\\
\notag\\
= &\mathbb{E}\left\{ {\left[ {\left( {{\bf{N}}_1^{\rm{u}}\left( {{l_1},{l_1}} \right) + {\bf{N}}_2^{\rm{u}}\left( {{l_1},{l_1}} \right)} \right)} \right.} \right.\notag\\
 \notag\\
 &\left. { + {{\left( {{\bf{N}}_1^{\rm{u}}\left( {{l_1},{l_1}} \right) + {\bf{N}}_2^{\rm{u}}\left( {{l_1},{l_1}} \right)} \right)}^*}} \right]\notag\\
\notag\\
   &\times \left[ {\left( {{\bf{N}}_1^{\rm{u}}\left( {{l_2},{l_2}} \right) + {\bf{N}}_2^{\rm{u}}\left( {{l_2},{l_2}} \right)} \right)} \right.\notag\\
\notag\\
 &\left. { + \left. {{{\left( {{\bf{N}}_1^{\rm{u}}\left( {{l_2},{l_2}} \right) + {\bf{N}}_2^{\rm{u}}\left( {{l_2},{l_2}} \right)} \right)}^*}} \right]} \right\}, \label{eq_48}
\end{align}
where ${\bf{N}}_1^{\rm{u}}\left( {{l_1},{l_1}} \right)$, ${\bf{N}}_1^{\rm{u}}\left( {{l_2},{l_2}} \right)$, ${\bf{N}}_2^{\rm{u}}\left( {{l_1},{l_1}} \right)$, and ${\bf{N}}_1^{\rm{u}}\left( {{l_2},{l_2}} \right)$ denote the ${\left( {{l_1},{l_1}} \right)}$-th and ${\left( {{l_2},{l_2}} \right)}$-th elements of the matrix ${\bf{N}}_1^{\rm{u}}$ and ${\bf{N}}_1^{\rm{u}}$.
In~\eqref{eq_48}, ${\bf{N}}_1^{\rm{u}}$ and ${\bf{N}}_2^{\rm{u}}$ is given by
\begin{align}
{\bf{N}}_1^{\rm{u}} =& \frac{1}{{{\sigma ^2}}}{{\bf{C}}^{\mathsf T}}{\left( {{\bf{Y}}_{(3)}^{\mathsf T} - \left( {{\bf{B}} \odot {\bf{A}}} \right){{\bf{C}}^{\mathsf T}}} \right)^{\mathsf H}}{{\bf{C}}^{\rm {a}}}\notag\\[6pt]
 = &\frac{1}{{{\sigma ^2}}}{{\bf{C}}^{\mathsf T}}{\left( {{{\bf{W}}_{(3)}}} \right)^*}{{\bf{C}}^{\rm {a}}}. \label{eq_49} \\[6pt]
{\bf{N}}_2^{\rm{u}} =& \frac{1}{{{\sigma ^2}}}{{\bf{C}}^{\rm {b}}}{\left( {{\bf{Y}}_{(3)}^{\mathsf T} - \left( {{\bf{B}} \odot {\bf{A}}} \right){{\bf{C}}^{\mathsf T}}} \right)^{\mathsf H}}\left( {{\bf{B}} \odot {\bf{A}}} \right)\notag\\[6pt]
 =& \frac{1}{{{\sigma ^2}}}{{\bf{C}}^{\rm {b}}}{\left( {{{\bf{W}}_{(3)}}} \right)^*}\left( {{\bf{B}} \odot {\bf{A}}} \right). \label{eq_50}
\end{align}
Vectorizing \eqref{eq_49} and \eqref{eq_50} yields
\begin{align}
{\bf{n}}_1^{\rm{u}} =& \frac{1}{{{\sigma ^2}}}\left( {{{\left( {{{\bf{C}}^{\rm {a}}}} \right)}^{\mathsf T}} \otimes {{\bf{C}}^{\mathsf T}}} \right){\rm vec}\left( {{{\left( {{{\bf{W}}_{(3)}}} \right)}^*}} \right),
\label{eq_53}\\
{\bf{n}}_2^{\rm{u}} =& \frac{1}{{{\sigma ^2}}}\left( {{{\left( {{\bf{B}} \odot {\bf{A}}} \right)}^{\mathsf T}} \otimes {{\bf{C}}^{\rm {b}}}} \right){\rm vec}\left( {{{\left( {{{\bf{W}}_{(3)}}} \right)}^*}} \right).\label{eq_54}
\end{align}

According to \eqref{eq_13} and \eqref{eq_30}, ${{\bf{W}}_{(1)}}$ is the model-1 expansion of the noise tensor. Since the noise obeys a complex Gaussian distribution, ${\rm vec}\left( {{\bf{W}}_{(1)}^*} \right)$ represents a zero-mean circular Gaussian random vector, which further yields that ${\bf{n}}_1^{\rm{u}}$ and ${\bf{n}}_2^{\rm{u}}$ also satisfy the symmetric complex circular Gaussian distribution. The covariance matrices associated with \eqref{eq_53} and \eqref{eq_54} can be written as
\begin{align}
{{\bf{C}}_{{\rm{n}}_1^{\rm{u}}}} =& \mathbb{E}\left\{ {\left( {{\bf{n}}_1^{\rm{u}}} \right){{\left( {{\bf{n}}_1^{\rm{u}}} \right)}^{\mathsf H}}} \right\}\notag\\[6pt]
  = &\frac{1}{{{\sigma ^2}}}\left( {{{\left( {{{\bf{C}}^{\rm {a}}}} \right)}^{\mathsf T}} \otimes {{\bf{C}}^{\mathsf T}}} \right)\left( {{{\left( {{{\bf{C}}^{\rm {a}}}} \right)}^*} \otimes {{\bf{C}}^*}} \right),\\
{{\bf{C}}_{{\rm{n}}_2^{\rm{u}}}} =& \mathbb{E}\left\{ {\left( {{\bf{n}}_2^{\rm{u}}} \right){{\left( {{\bf{n}}_2^{\rm{u}}} \right)}^{\mathsf H}}} \right\}\notag\\[6pt]
  = &\frac{1}{{{\sigma ^2}}}\left( {{{\left( {{\bf{B}} \odot {\bf{A}}} \right)}^{\mathsf T}} \otimes {{\bf{C}}^{\rm {b}}}} \right){\left( {{{\left( {{\bf{B}} \odot {\bf{A}}} \right)}^{\mathsf T}} \otimes {{\bf{C}}^{\rm {b}}}} \right)^{\mathsf H}},\\
{{\bf{C}}_{{\rm{n}}_1^{\rm{u}},{\rm{n}}_2^{\rm{u}}}} =& \mathbb{E}\left\{ {\left( {{\bf{n}}_1^{\rm{u}}} \right){{\left( {{\bf{n}}_2^{\rm{u}}} \right)}^{\mathsf H}}} \right\}\notag\\[6pt]
  = &\frac{1}{{{\sigma ^2}}}\left( {{{\left( {{{\bf{C}}^{\rm {a}}}} \right)}^{\mathsf T}} \otimes {{\bf{C}}^{\mathsf T}}} \right){\left( {{{\left( {{\bf{B}} \odot {\bf{A}}} \right)}^{\mathsf T}} \otimes {{\bf{C}}^{\rm {b}}}} \right)^{\mathsf H}},\\
{{\bf{C}}_{{\rm{n}}_2^{\rm{u}},{\rm{n}}_1^{\rm{u}}}} =& \mathbb{E}\left\{ {\left( {{\bf{n}}_2^{\rm{u}}} \right){{\left( {{\bf{n}}_1^{\rm{u}}} \right)}^{\mathsf H}}} \right\}\notag\\[6pt]
  =& \frac{1}{{{\sigma ^2}}}\left( {{{\left( {{\bf{B}} \odot {\bf{A}}} \right)}^{\mathsf T}} \otimes {{\bf{C}}^{\rm {b}}}} \right){\left( {{{\left( {{{\bf{C}}^{\rm {a}}}} \right)}^{\mathsf T}} \otimes {{\bf{C}}^{\mathsf T}}} \right)^{\mathsf H}}.
\end{align}
The result of \eqref{eq_48} can then be written as
\begin{align}
&{{\bf{J}}_{44}}\left( {{l_1},{l_2}} \right)=\mathbb{E}\left\{ {{{\left( {\frac{{\partial L_{{\rm{o}}}\left( {{\bf{p}}_{\rm{e}}} \right)}}{{\partial {\tau_{{l_1}}}}}} \right)}^*}\left( {\frac{{\partial L_{{\rm{o}}}\left( {{\bf{p}}_{\rm{e}}} \right)}}{{\partial {\tau_{{l_2}}}}}} \right)} \right\}\notag\\[8pt]
 =& 2{\mathop{\rm Re}\nolimits} \left\{ {{{\bf{C}}_{{\rm{n}}_1^{\rm{u}}}}\left( {m,n} \right)} \right\} + 2{\mathop{\rm Re}\nolimits} \left\{ {{{\bf{C}}_{{\rm{n}}_2^{\rm{u}}}}\left( {m,n} \right)} \right\}\notag\\[8pt]
+ & 2{\mathop{\rm Re}\nolimits} \left\{ {{{\bf{C}}_{{\rm{n}}_1^{\rm{u}},{\rm{n}}_2^{\rm{u}}}}\left( {m,n} \right)} \right\} + 2{\mathop{\rm Re}\nolimits} \left\{ {{{\bf{C}}_{{\rm{n}}_2^{\rm{u}},{\rm{n}}_1^{\rm{u}}}}\left( {m,n} \right)} \right\},
\end{align}
where $m = {L}\left( {{l_1} - 1} \right) + {\kern 1pt} {l_1}$ and $n = {L}\left( {{l_2} - 1} \right) + {\kern 1pt} {l_2}$.

Similarly, the main diagonal submatrix of the FIM with respect to ${\theta _{{\rm{e}}}}$ is obtained as
\begin{align}
{{\bf{J}}_{11}}\left( {{l_1},{l_2}} \right)=&\mathbb{E}\left\{ {{{\left( {\frac{{\partial L_{{\rm{o}}}\left( {{\bf{p}}_{\rm{e}}} \right)}}{{\partial {\theta _{{\rm{e}},{l_1}}}}}} \right)}^*}\left( {\frac{{\partial L_{{\rm{o}}}\left( {{\bf{p}}_{\rm{e}}} \right)}}{{\partial {\theta _{{\rm{e}},{l_2}}}}}} \right)} \right\}\notag\\[8pt]
=& 2{\rm{Re}}\left\{ {{{\bf{C}}_{{{\rm{n^a}}}}}\left( {m,n} \right)} \right\},
\end{align}
where the covariance matrix is ${{\bf{C}}_{{{{\rm{n}^a}}}}} = \frac{1}{{{\sigma ^2}}}\left( {{\bf{\tilde A}}_1^{\mathsf T} \otimes {{\left( {{\bf{C}} \odot {\bf{B}}} \right)}^{\mathsf T}}} \right)\left( {{\bf{\tilde A}}_1^* \otimes {{\left( {{\bf{C}} \odot {\bf{B}}} \right)}^*}} \right)$, and  the vectorized noise-related term is denoted as
\begin{equation}
{{\bf{n}}^{\rm{a}}} = \frac{1}{{{\sigma ^2}}}\left( {{\bf{\tilde A}}_1^{\mathsf T} \otimes {{\left( {{\bf{C}} \odot {\bf{B}}} \right)}^{\mathsf T}}} \right){\rm vec}\left( {\left( {{\bf{W}}_{(1)}^{\mathsf H}} \right)} \right).
\label{eq_61}
\end{equation}

For the parameter ${ \phi _{{\rm{a}}}}$, we have
\begin{align}
{{\bf{J}}_{22}}\left( {{l_1},{l_2}} \right)=&\mathbb{E}\left\{ {{{\left( {\frac{{\partial L_{{\rm{o}}}\left( {{\bf{p}}_{\rm{e}}} \right)}}{{\partial { \phi _{{\rm{a}},{l_1}}}}}} \right)}^*}\left( {\frac{{\partial L_{{\rm{o}}}\left( {{\bf{p}}_{\rm{e}}} \right)}}{{\partial { \phi _{{\rm{a}},{l_2}}}}}} \right)} \right\} \notag\\[8pt]
=& 2{\mathop{\rm Re}\nolimits} \left\{ {{{\bf{C}}_{{\rm{n^b}}}}\left( {m,n} \right)} \right\},
\end{align}
where ${{\bf{C}}_{{\rm{n^b}}}} = \frac{1}{{{\sigma ^2}}}\left( {{\bf{\tilde A}}_2^{\mathsf T} \otimes {{\left( {{\bf{C}} \odot {\bf{B}}} \right)}^{\mathsf T}}} \right)\left( {{\bf{\tilde A}}_2^* \otimes {{\left( {{\bf{C}} \odot {\bf{B}}} \right)}^*}} \right)$, and 
\begin{equation}
{{\bf{n}}^{\rm{b}}} = \frac{1}{{{\sigma ^2}}}\left( {{\bf{\tilde A}}_2^{\mathsf T} \otimes {{\left( {{\bf{C}} \odot {\bf{B}}} \right)}^{\mathsf T}}} \right){\rm vec}\left( {\left( {{\bf{W}}_{(1)}^{\mathsf H}} \right)} \right).
\label{eq_63}
\end{equation}

For the parameter ${{\psi }}$, we have
\begin{align}
{{\bf{J}}_{33}}\left( {{l_1},{l_2}} \right)=&\mathbb{E}\left\{ {{{\left( {\frac{{\partial L_{{\rm{o}}}\left( {{\bf{p}}_{\rm{e}}} \right)}}{{\partial {{\psi _{l_1}}}}}} \right)}^*}\left( {\frac{{\partial L_{{\rm{o}}}\left( {{\bf{p}}_{\rm{e}}} \right)}}{{\partial {{\psi _{l_2}}}}}} \right)} \right\} \notag\\[8pt]
= & 2{\mathop{\rm Re}\nolimits} \left\{ {{{\bf{C}}_{{\rm{n^c}}}}\left( {m,n} \right)} \right\},
\end{align}
where ${{\bf{C}}_{{\rm{n^c}}}} = \frac{1}{{{\sigma ^2}}}\left( {{{{\bf{\tilde B}}}^{\mathsf T}} \otimes {{\left( {{\bf{C}} \odot {\bf{A}}} \right)}^{\mathsf T}}} \right)\left( {{{{\bf{\tilde B}}}^*} \otimes {{\left( {{\bf{C}} \odot {\bf{A}}} \right)}^*}} \right)$, and
\begin{equation}
{{\bf{n}}^{\rm{c}}} = \frac{1}{{{\sigma ^2}}}\left( {{\bf{\tilde A}}_2^{\mathsf T} \otimes {{\left( {{\bf{C}} \odot {\bf{B}}} \right)}^{\mathsf T}}} \right){\rm vec}\left( {\left( {{\bf{W}}_{(2)}^{\mathsf H}} \right)} \right).
\label{eq_65}
\end{equation}

For the parameter ${\gamma }$, it can be written as
\begin{align}
{{\bf{J}}_{55}}\left( {{l_1},{l_2}} \right)=& \mathbb{E}\left\{ {{{\left( {\frac{{\partial L_{{\rm{o}}}\left( {{\bf{p}}_{\rm{e}}} \right)}}{{\partial {\gamma _{{l_1}}}}}} \right)}^*}\left( {\frac{{\partial L_{{\rm{o}}}\left( {{\bf{p}}_{\rm{e}}} \right)}}{{\partial {\gamma _{{l_2}}}}}} \right)} \right\} \notag\\[8pt]
=&{{\bf{C}}_{{\rm{n^g}}}}{\left( {m,n} \right)^*},
\end{align}
\\
where ${{\bf{C}}_{{\rm{n^g}}}} = \frac{1}{{{\sigma ^2}}}\left( {{{{\bf{\tilde G}}}^{\mathsf T}} \otimes {{\left( {{\bf{B}} \odot {\bf{A}}} \right)}^{\mathsf T}}} \right)\left( {{{{\bf{\tilde G}}}^*} \otimes {{\left( {{\bf{B}} \odot {\bf{A}}} \right)}^*}} \right),$ and 
\begin{equation}
{{\bf{n}}^{\rm{g}}} = \frac{1}{{{\sigma ^2}}}\left( {{{{\bf{\tilde G}}}^{\mathsf T}} \otimes {{\left( {{\bf{B}} \odot {\bf{A}}} \right)}^{\mathsf T}}} \right){\rm vec}\left( {\left( {{\bf{W}}_{(3)}^{\mathsf H}} \right)} \right).
\label{eq_67}
\end{equation}

Subsequently, the nondiagonal submatrices of the FIM are obtained. The submatrices associated with parameters ${\tau}$ and ${\theta}$ can be written as 
\begin{align}
&{{\bf{J}}_{41}}\left( {{l_1},{l_2}} \right)=\mathbb{E}\left\{ {{{\left( {\frac{{\partial L_{{\rm{o}}}\left( {{\bf{p}}_{\rm{e}}} \right)}}{{\partial {\tau_{{l_1}}}}}} \right)}^*}\left( {\frac{{\partial L_{{\rm{o}}}\left( {{\bf{p}}_{\rm{e}}} \right)}}{{\partial {\theta _{{\rm{e}},{l_2}}}}}} \right)} \right\}\notag\\[8pt]
= &4\mathbb{E}\left\{ {{\mathop{\rm Re}\nolimits} \left( {{\bf{e}}_{{l_1}}^{\mathsf T}{\bf{N}}_1^{\rm{u}}{{\bf{e}}_{{l_1}}} + {\bf{e}}_{{l_1}}^{\mathsf T}{\bf{N}}_2^{\rm{u}}{{\bf{e}}_{{l_1}}}} \right){\mathop{\rm Re}\nolimits} \left( {{\bf{e}}_{{l_2}}^{\mathsf T}{{\bf{N}}^{\rm{a}}}{{\bf{e}}_{{l_2}}}} \right)} \right\}\notag\\[8pt]
= &\mathbb{E}\left\{ {\left( {\left[ {{\bf{N}}_1^{\rm{u}}\left( {{l_1},{l_1}} \right) + {\bf{N}}_2^{\rm{u}}\left( {{l_1},{l_1}} \right)} \right]} \right.} \right.\notag\\[8pt]
&+ \left. {{{\left[ {{\bf{N}}_1^{\rm{u}}\left( {{l_1},{l_1}} \right) + {\bf{N}}_2^{\rm{u}}\left( {{l_1},{l_1}} \right)} \right]}^*}} \right)\notag\\[8pt]
&\times \left. {\left( {{{\bf{N}}^{\rm{a}}}\left( {{l_2},{l_2}} \right) + {{\bf{N}}^{\rm{a}}}{{\left( {{l_2},{l_2}} \right)}^*}} \right)} \right\}.
\label{eq_68}
\end{align}

To derive the closed-form expression of \eqref{eq_68}, recalling \eqref{eq_53} and \eqref{eq_54} yields
\begin{align}
&{{\bf{C}}_{{\rm{n_1^u}},{\rm{n^a}}}} =\mathbb{E}\left\{ {\left( {{\bf{n}}_1^u} \right){{\left( {{{\bf{n}}^a}} \right)}^{\mathsf H}}} \right\}\notag\\[8pt]
&\quad\quad \ \  = \frac{1}{{{\sigma ^4}}}\left( {{{\left( {{{\bf{C}}^{\rm {a}}}} \right)}^{\mathsf T}} \otimes {{\bf{C}}^{\mathsf T}}} \right){{\bf{C}}_{3,1}}{\left( {{\bf{\tilde A}}_1^{\mathsf T} \otimes {{\left( {{\bf{C}} \odot {\bf{B}}} \right)}^{\mathsf T}}} \right)^{\mathsf H}}, \\
&{{\bf{C}}_{{\rm{n_2^u}},{{\rm{n^a}}}}} = \mathbb{E}\left\{ {\left( {{\bf{n}}_2^u} \right){{\left( {{{\bf{n}}^a}} \right)}^{\mathsf H}}} \right\}\notag\\
&\quad\quad \ \ = \frac{1}{{{\sigma ^4}}}\left( {{{\left( {{\bf{B}} \odot {\bf{A}}} \right)}^{\mathsf T}} \otimes {{\bf{C}}^{\rm {b}}}} \right){{\bf{C}}_{3,1}}{\left( {{\bf{\tilde A}}_1^{\mathsf T} \otimes {{\left( {{\bf{C}} \odot {\bf{B}}} \right)}^{\mathsf T}}} \right)^{\mathsf H}},
\end{align}
where ${{\bf{C}}_{3,1}} = \mathbb{E}\left\{ {{\rm vec}\left( {{\bf{W}}_{(3)}^*} \right){\rm vec}\left( {{{\bf{W}}_{(1)}}} \right)} \right\}$. Finally, 
we obtain the result of \eqref{eq_68} as
\begin{align}
&{{\bf{J}}_{41}}\left( {{l_1},{l_2}} \right)=\mathbb{E}\left\{ {{{\left( {\frac{{\partial L_{{\rm{o}}}\left( {{\bf{p}}_{\rm{e}}} \right)}}{{\partial {\tau_{{l_1}}}}}} \right)}^*}\left( {\frac{{\partial L_{{\rm{o}}}\left( {{\bf{p}}_{\rm{e}}} \right)}}{{\partial {\theta _{{\rm{e}},{l_2}}}}}} \right)} \right\}\notag\\[8pt]
= &2{\mathop{\rm Re}\nolimits} \left\{ {{{\bf{C}}_{{\rm{n_1^u}},{{\rm{n^a}}}}}\left( {m,n} \right)} \right\}+ 2{\mathop{\rm Re}\nolimits} \left\{ {{{\bf{C}}_{{\rm{n_2^u}},{{\rm{n^a}}}}}\left( {m,n} \right)} \right\}.
\label{eq_70}
\end{align}

We next derive the value of the FIM element independent of the time delay. The submatrices with respect to the parameters ${\theta}$ and ${\phi}$ are given by
\begin{align}
{{\bf{J}}_{12}}\left( {{l_1},{l_2}} \right)=&{\rm{E}}\left\{ {{{\left( {\frac{{\partial L_{{\rm{o}}}\left( {{\bf{p}}_{\rm{e}}} \right)}}{{\partial {\theta _{{\rm{e}},{l_1}}}}}} \right)}^*}\left( {\frac{{\partial L_{{\rm{o}}}\left( {{\bf{p}}_{\rm{e}}} \right)}}{{\partial {\phi _{{\rm{a}},{l_2}}}}}} \right)} \right\}\notag\\[8pt]
 = &4\mathbb{E}\left\{ {{\mathop{\rm Re}\nolimits} \left( {{\bf{e}}_{{l_1}}^{\mathsf T}{{\bf{N}}^{\rm{a}}}{{\bf{e}}_{{l_1}}}} \right){\mathop{\rm Re}\nolimits} \left( {{\bf{e}}_{{l_2}}^{\mathsf T}{{\bf{N}}^{\rm{c}}}{{\bf{e}}_{{l_2}}}} \right)} \right\}\notag\\[8pt]
 = &2{\mathop{\rm Re}\nolimits} \left\{ {{{\bf{C}}_{{\rm{n^a}},{{\rm{n^c}}}}}\left( {m,n} \right)} \right\},
 \label{eq_71}
\end{align}
where ${{\bf{C}}_{{\rm{n^a}},{{\rm{n^c}}}}} = \frac{1}{{{\sigma ^4}}}\left( {{\bf{\tilde A}}_1^{\mathsf T} \otimes {{\left( {{\bf{C}} \odot {\bf{B}}} \right)}^{\mathsf T}}} \right){{\bf{C}}_{1,3}}\left( {{{{\bf{\tilde B}}}^*} \otimes {{\left( {{\bf{C}} \odot {\bf{A}}} \right)}^*}} \right),$ ${{\bf{C}}_{1,3}} = \mathbb{E}\left\{ {{\rm vec}\left( {{\bf{W}}_{(1)}^{\mathsf H}} \right){\rm vec}\left( {{{\bf{W}}_{(1)}}} \right)} \right\}.$

The analysis process of the expression for the remaining nondiagonal submatrices of the FIM matrix is similar to the analysis process of \eqref{eq_70} and \eqref{eq_71}.

The dimension of the noise tensor is $Q \times P \times T_{{\rm{a}}} $. Here, we use $\left( {q,p,t} \right)$ to denote the index of each dimension of the noise tensor. The $\left( {q,p,t} \right)$-th element of the noise tensor can be respectively represented the values of its model-1, model-2, and model-3 expansions, as given by
\begin{align}
&{{\bf{W}}_{(1)}}\left\{ {q,\left( {p - 1} \right)T_{{\rm{a}}}  + t} \right\}\notag\\[8pt]
 &\to {\rm vec}\left( {{\bf{W}}_{_{(1)}}^{\mathsf H}} \right)\left\{ {\left( {p - 1} \right)T_{{\rm{a}}}  + t + \left( {q - 1} \right)PT_{{\rm{a}}} } \right\},
 \label{eq_72}
\end{align}
\begin{align}
&{{\bf{W}}_{(2)}}\left\{ {t,\left( {p - 1} \right)Q + q} \right\}\notag\\[8pt]
& \to {\rm vec}\left( {{\bf{W}}_{_{(2)}}^{\mathsf H}} \right)\left\{ {\left( {p - 1} \right)Q + q + \left( {t - 1} \right)PQ} \right\},
\label{eq_73}
\end{align}
\begin{align}
&{{\bf{W}}_{(3)}}\left\{ {p,\left( {t - 1} \right)Q + q} \right\}\notag\\[8pt]
 &\to {\rm vec}\left( {{\bf{W}}_{_{(3)}}^{\mathsf H}} \right)\left\{ {\left( {t - 1} \right)Q + q + \left( {p - 1} \right)QT_{{\rm{a}}} } \right\}.
 \label{eq_74}
\end{align}

Since each element of the noise tensor is i.i.d., the mean of any two of its modes has a value only if the indices are equal, and zero otherwise. For example, ${{\bf{C}}_{1,3}}$ can be written as
\begin{small} 
\begin{equation}
    \mathbb{E}\left\{ {{w_{{q_1},{t_1},{p_1}}}w_{{q_2},{t_2},{p_2}}^*} \right\}=
    \begin{cases}
        {\sigma ^2}; \!{q_1} \!= \!{q_2},\!{t_1} \!= \!{t_{2,}}{p_1} \!= \!{p_2}\\[8pt]
        0; \;\;\; \mathrm{otherwise} 
    \end{cases}
    \label{eq_75}
\end{equation}
\end{small}
From \eqref{eq_75}, there are $QT_{{\rm{a}}} P$ nonzero elements of ${{\bf{C}}_{1,3}}$, and their indices are $\left\{ {\left( {p - 1} \right)T_{{\rm{a}}}  + t + \left( {q - 1} \right)PT_{{\rm{a}}} } \right.
\left. {,\left( {p - 1} \right)Q + q + \left( {t - 1} \right)PQ} \right\}$.

Based on \eqref{eq_72}, \eqref{eq_73}, \eqref{eq_74}, and \eqref{eq_75}, we can calculate the specific value of each element of the FIM matrix.

After calculating the CRLB for each parameter, we next calculate the CRLB for the channel  consisting of each parameter. The one in \eqref{eq_1} represents the channel of the $p$-th subcarrier. After considering all the time slots, time frames, and subcarriers, the channel can be represented in tensor form as ${{\boldsymbol{{\cal H}}}_{\rm{u}}}$. According to \cite{r34}, the tensor channel model can be decomposed into tensor $n$-mode products, which can be written as 
\begin{equation}
{{\boldsymbol{{\cal H}}}_{\rm{u}}} = {{\boldsymbol{{\cal I}}}_{4,L,}}{ \times _1}{{\bf{A}}_{\rm{R}}}{ \times _2}{\bf{A}}_{\rm{U}}^*{ \times _3}{{\bf{C}}}{ \times _4}{{\bm{\upgamma }}^{{T}}},
\end{equation}
where ${ \times _n}$ represents the tensor's $n$-mode product, $\left (\cdot \right ) ^*$ is the conjugate, $L$, ${\bm{\gamma }}$ and ${{\bf{C}}}$  have been defined previously, and ${{\boldsymbol{{\cal I}}}_{4,L,}}$ denotes a four-dimensional superdiagonal tensor with superdiagonal positions of $1$ and other positions of $0$ \cite{r35}. According to \eqref{eq_3} and \eqref{eq_4}, ${{\bf{A}}_{\rm{R}}}$ and ${\bf{A}}_{\rm{U}}$ can be written as  
\begin{align}
{{\bf{A}}_{\rm{R}}} &= \left[ {{\bf{ a}}_{{\rm{IRS}}}\left( {{\theta _{{\rm{e}},1}},{\phi _{{\rm{a}},1}},{u_1}} \right), \ldots ,{\bf{ a}}_{{\rm{IRS}}}\left( {{\theta _{{\rm{e}},L}},{\phi _{{\rm{a}},L}},{u_L}} \right)} \right],\notag\\[5pt]
{\bf{A}}_{\rm{U}} &= \left[ {{{{\bf{a}}}_{{\rm{ue}}}}\left( {{\psi _1}} \right), \ldots ,{{{\bf{a}}}_{{\rm{ue}}}}\left( {{\psi _L}} \right)} \right],
\end{align}

Vectorizing the tensor ${{\bf{{\cal H}}}_{\rm{u}}}$, we have 
\begin{align}
{\bf{h}}_{\rm{v}}=({\bf{C}} \odot {\bf{A}}_\mathrm{U}^*\odot  {\bf{A}}_\mathrm{R}){\bm{\upgamma}},
\label{eq_79}
\end{align}
where ${\bf{h}}_{\rm{v}} = {\rm{vec}}\left\{ {{\boldsymbol{{\cal H}}}_{\rm{u}}} \right\}$, and ${\rm{vec}}\left\{ \cdot  \right\}$ denote the vectorization operation.

According to \cite{r36}, the CRLB of the overall channel can be expressed as 
\begin{align}
\mathrm{CRLB}({\bf{h}}_{\rm{v}})=\mathrm{tr}\left[\left(\frac{\partial {\bf{h}}_{\rm{v}}}{\partial\left({{\bf{p}}_{\rm{e}}}\right)^\mathrm{T}}\right)\mathbf{J}^{-1}{\left({{\bf{p}}_{\rm{e}}}\right)}\left(\frac{\partial{\bf{h}}_{\rm{v}}}{\partial\left({{\bf{p}}_{\rm{e}}}\right)^\mathrm{T}}\right)^{\mathsf H}\right],
\label{eq_80}
\end{align}
where
\begin{align}
\frac{\partial{\bf{h}}_{\rm{v}}}{\partial {\bf p}_\mathrm{e}}=&\left[\frac{\partial{\bf{h}}_{\rm{v}}}{\partial\theta_\mathrm{e,1}},...,\frac{\partial{\bf{h}}_{\rm{v}}}{\partial\theta_{\mathrm{e},L}},\frac{\partial{\bf{h}}_{\rm{v}}}{\partial\phi_\mathrm{a,1}},...,\frac{\partial{\bf{h}}_{\rm{v}}}{\partial\phi_{\mathrm{a},L}},\frac{\partial{\bf{h}}_{\rm{v}}}{\partial\psi_1},...,\frac{\partial{\bf{h}}_{\rm{v}}}{\partial\psi_L},
\right.\notag\\[8pt]&\left.
\quad\frac{\partial{\bf{h}}_{\rm{v}}}{\partial\tau_1},...,\frac{\partial{\bf{h}}_{\rm{v}}}{\partial\tau_L},\frac{\partial{\bf{h}}_{\rm{v}}}{\partial\gamma_1},...,\frac{\partial{\bf{h}}_{\rm{v}}}{\partial\gamma_L}\right].
\label{eq_81}
\end{align}

Given the similarity in computing each element of \eqref{eq_81}, we will focus solely on deriving the first element of \eqref{eq_81} as a representative example. Based on \eqref{eq_79}, we have 
\begin{align}
\frac{\partial{\bf{h}}_{\rm{v}}}{\partial\theta_\mathrm{e,1}}=&\frac{\partial{(\mathbf{C}\odot {\bf{A}}_\mathrm{U}^*\odot {\bf{A}}_\mathrm{R}){\bm{\upgamma}}}}{\partial\theta_\mathrm{e,1}}
\notag\\[8pt]
=& \left(\mathbf{C}\odot {\bf{A}}_\mathrm{U}^*\odot\frac{\partial {\bf{A}}_\mathrm{R}}{\partial \theta_\mathrm{e,1}} \right){\bm{\upgamma}},
\end{align}
where the derivation of $\frac{\partial {\bf{A}}_\mathrm{R}}{\partial \theta_\mathrm{e,1}}$ follows the same principle as that used in \eqref{eq_40}. The derivation for the other elements proceeds analogously.
Finally, the CRLB for the overall channel can be obtained by substituting the derivative of each element into \eqref{eq_80}.

\bibliographystyle{IEEEtran}
\bibliography{myref}

\begin{IEEEbiography}[{\includegraphics[width=1in,height=1.25in,clip,keepaspectratio]{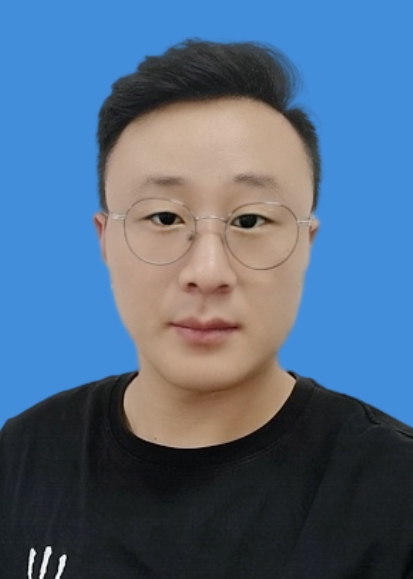}}] {Wenzhou Cao} received the B.S. and M.S. degrees in electronic and information engineering from the Zhongyuan University of Technology (ZUT), Zhengzhou, China, in 2018 and 2021, respectively. He is currently pursuing the Ph.D. degree in the School of Information and Communication Engineering, Beijing University of Posts and Telecommunications (BUPT), Beijing, China. His research interests include intelligent reflecting surface (IRS), channel estimation, and tensor decomposition.
\end{IEEEbiography}

\begin{IEEEbiography}[{\includegraphics[width=1in,height=1.25in,clip,keepaspectratio]{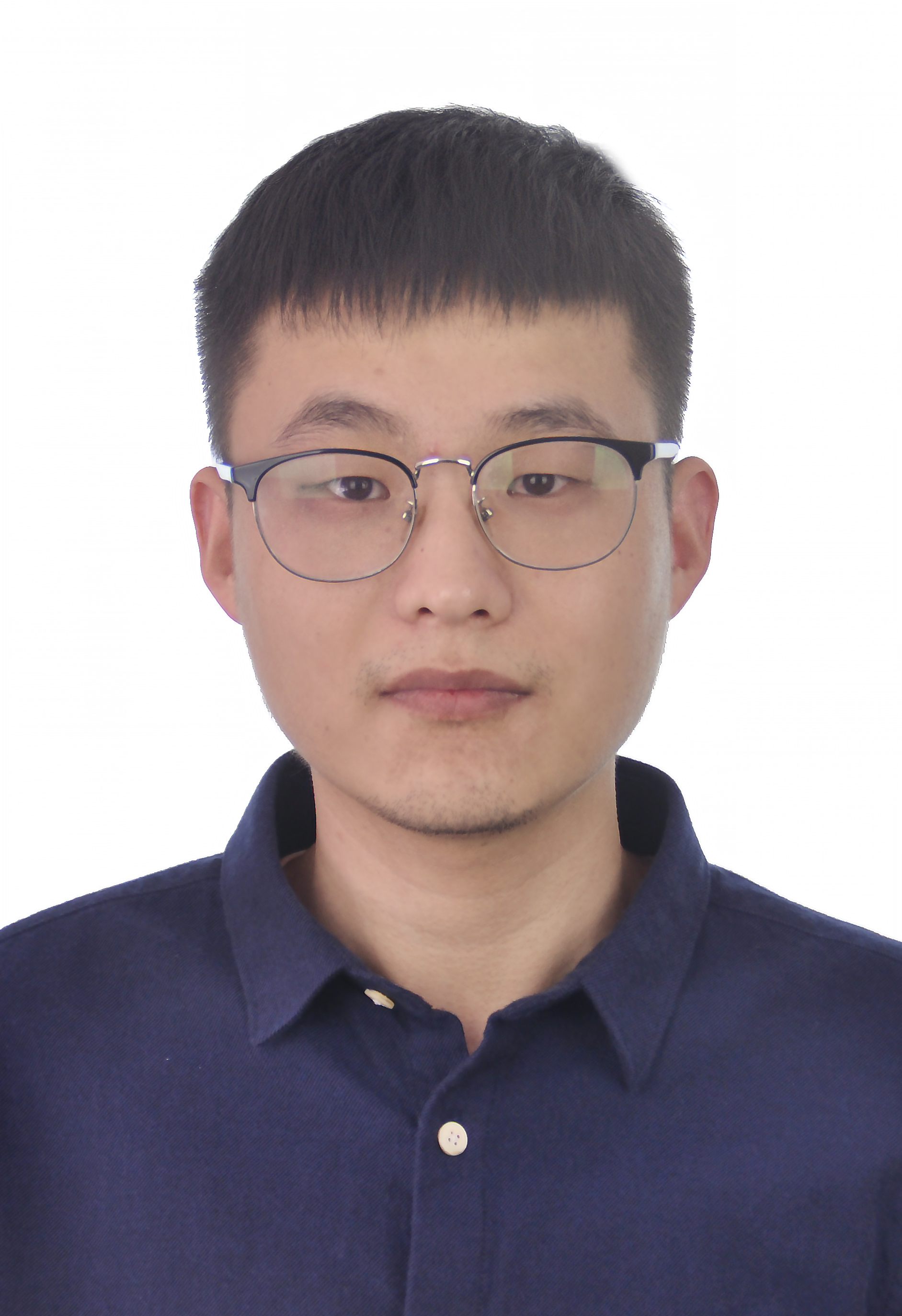}}]{Yashuai Cao} received the B.E. and Ph.D. degrees in communication engineering from Chongqing University of Posts and Telecommunications (CQUPT) and Beijing University of Posts and Telecommunications (BUPT), China, in 2017 and 2022, respectively. From 2022 to 2023, he was a lecturer in the Department of Electronics and Communication Engineering, North China Electric Power University (NCEPU), Baoding. From 2023 to 2025, he was a Postdoctoral Research Fellow with the Department of Electronic Engineering, Tsinghua University, Beijing, China. He is currently a Distinguished Associate Professor with the School of Intelligence Science and Technology, University of Science and Technology Beijing (USTB), Beijing, China. His research interests include Stacked Intelligent Metasurface, Environment-Aware Communications, and Channel Knowledge Map.
\end{IEEEbiography}

\begin{IEEEbiography}[{\includegraphics[width=1in,height=1.25in,clip,keepaspectratio]{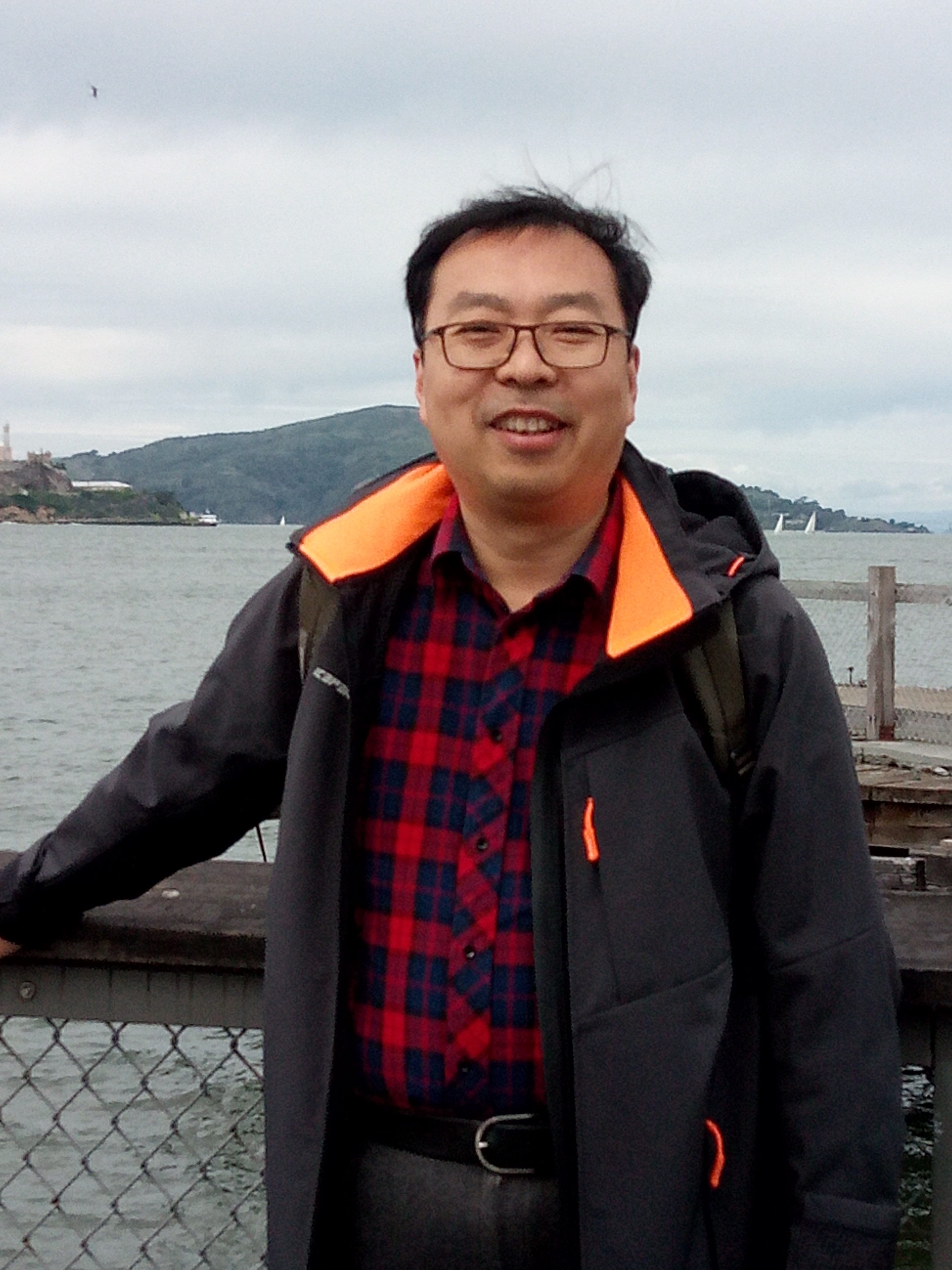}}]{Tiejun Lv} (Senior Member, IEEE) received the M.S. and Ph.D. degrees in electronic engineering from the University of Electronic Science and Technology of China (UESTC), Chengdu, China, in 1997 and 2000, respectively. From January 2001 to January 2003, he was a Post-Doctoral Fellow at Tsinghua University, Beijing, China. In 2005, he was promoted to a Full Professor at the School of Information and Communication Engineering, Beijing University of Posts and Telecommunications (BUPT). From September 2008 to March 2009, he was a Visiting Professor with the Department of Electrical Engineering, Stanford University, Stanford, CA, USA. He is currently the author of four books, one book chapter, and more than 170 published journal articles and 200 conference papers on the physical layer of wireless mobile communications. His current research interests include signal processing, communications theory, and networking. He was a recipient of the Program for New Century Excellent Talents in University Award from the Ministry of Education, China, in 2006. He received the Nature Science Award from the Ministry of Education of China for the hierarchical cooperative communication theory and technologies in 2015 and Shaanxi Higher Education Institutions Outstanding Scientific Research Achievement Award in 2025.
\end{IEEEbiography}

\begin{IEEEbiography}[{\includegraphics[width=1in,height=1.25in,clip,keepaspectratio]{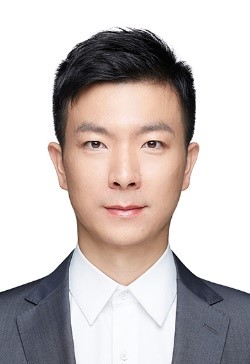}}]{Jie Zeng} received the B.S. and M.S. degrees from Tsinghua University in 2006 and 2009, respectively, and received two Ph.D. degrees from Beijing University of Posts and Telecommunications in 2019 and the University of Technology Sydney in 2021, respectively. From July 2009 to May 2020, he was with the Research Institute of Information Technology, Tsinghua University. From May 2020 to April 2022, he was a postdoctoral researcher with the Department of Electronic Engineering, Tsinghua University. Since May 2022, he has been an associate professor with the School of Cyberspace Science and Technology, Beijing Institute of Technology. His research interests include 5G/6G, URLLC, satellite Internet, and novel network architecture. He has published over 100 journal and conference papers, and holds more than 40 Chinese and international patents. He participated in drafting one national standard and one communication industry standard in China. 
He received Beijing’s science and technology award in 2015, the best cooperation award of Samsung Electronics in 2016, and Dolby Australia’s best scientific paper award in 2020.
\end{IEEEbiography}

\end{document}